# Magnetic Geometry and Physics of Advanced Divertors: The X-Divertor and the Snowflake

Mike Kotschenreuther, Prashant Valanju, Brent Covele, and Swadesh Mahajan

Institute for Fusion Studies, The University of Texas at Austin.

## Abstract

Advanced divertors are magnetic geometries where a second X-point is added in the divertor region to address the serious challenges of burning plasma power exhaust. Invoking physical arguments, numerical work, and detailed model magnetic field analysis, we investigate the magnetic field structure of advanced divertors in the physically relevant region for power exhaust – the Scrape-Off Layer (SOL). A primary result of our analysis is the emergence of a physical "metric", the Divertor Index *DI*, that quantifies the flux expansion increase as one goes from the main X-point to the strike point. It clearly separates three geometries with distinct consequences for divertor physics – the Standard Divertor  (SD, *DI = 1*), and two advanced geometries – the X-Divertor (XD, *DI > 1*) and the Snowflake (SFD, *DI < 1*). The XD, therefore, cannot be classified as one variant of the Snowflake. By this measure, recent NSTX and DIIID experiments are X-Divertors, not Snowflakes.

## I. Introduction

The problem of heat and particle exhaust in fusion grade plasmas is, currently, recognized as a very serious issue[1,2,3], serious enough that innovative approaches must be developed in the quest for fusion power. There is a greater and greater realization that the generic Standard Divertor  (SD), the class of magnetic exhaust configurations that are employed in the current machines (and also being planned for ITER), might be problematic for the high-performance era of ITER operation, and inadequate for a putative reactor.

Fortunately, the enormity of the exhaust problem was anticipated a while ago, and attempts to find appropriate solutions have led to the creation of novel magnetic geometries with two X-points[1,2,3,4,5,6]. Such geometries, collectively labeled here as Advanced Diverors (AD), include the "X-Divertor  (XD)" introduced in 2004[1,2,3,*], "Snowflake Divertors (SFD)" in 2007[4,5], the Super-X Divertor (SXD) in 2007[7,8,9], and "Asymmetric Snowflake Divertors" in 2010[6].

These conceptual developments stimulated robust experimental programs to create advanced geometries on current tokamaks (see Fig.1): TCV[10], NSTX[11,12] and DIIID[13] in 2009-2012. Experiments to realize the more ambitious SXD are scheduled for 2015 on the MAST tokamak[14].

One of the most pressing issues in the wake of these early experiments is to understand, in depth, the nature of the successful magnetic geometries (in contrast to what were termed "failure modes" in one NSTX publication[12]), and delineate the commonalities and differences in the observed and anticipated physics that these geometries may imply. Precise answers to these questions are essential in determining whether and how such (or similar) geometries would extrapolate to bigger and more powerful machines like the ITER and DEMO fusion reactors.

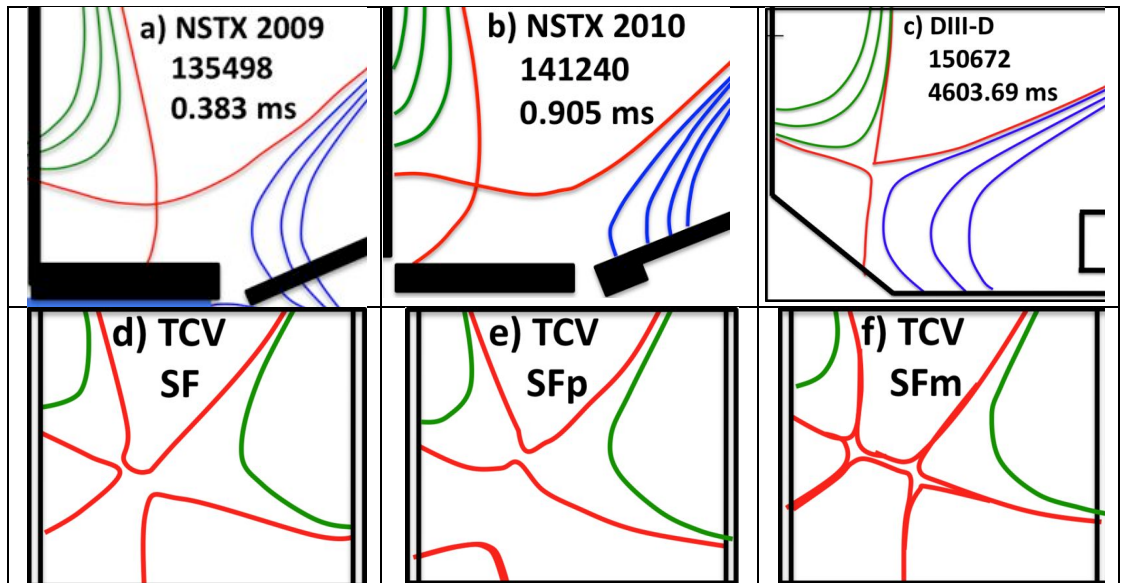

FIG. 1. SOL field lines from Advanced Divertor (AD) experiments. Cases showing flared field lines at the divertor plate (XD magnetic geometry) include a) NSTX (in 2009[11]), b) NSTX (in 2010[12]), c) DIII-D[13], while strongly convergent field lines at the divertor plate (SFD geometry) are seen in TCV[10] : d) nearly exact SFD, e) SFD plus, and f) SFD minus. The outer divertor plates are much further than the separation between the two X-points in TCV. The difference in the SOL geometry is quite apparent: in a)-c) the SOL expands as it approaches the strike point, while in d)-f) it expands near the main X-point.

Since experiments have successfully generated both the XD (NSTX and DIII-D) and the SFD (TCV), one would hope that such geometries can be created for future machines (superconducting and otherwise) including ITER. For ITER, however, one must work within severe constraints imposed by the fact that its coil set specifications are, by now, "set in stone". It may be of interest to the readers of this paper that we have been successful in designing X-Divertors for ITER, and tokamaks like ITER. We can accomplish this with reasonably low currents in superconducting poloidal coils that are all outside the toroidal coils. This is the subject of a forthcoming publication; the manuscript is under preparation.

For a deeper understanding of the advanced divertor geometries, a broad conceptual and theoretical framework is needed. One such framework was developed in Ref. 6 where a model analytical two X-point vacuum magnetic field was analyzed. The analysis of Ref. 6 described all two-X-point geometries as Snowflakes. Naming such a class implies that there is a shared property among the members of the class. The common, "Snowflake-like" property described in Ref. 6 was the similarity in the behavior of the vacuum magnetic field in a region that is frequently free of plasma (where the flux surfaces have three separate lobes). Ref. 6 extended the earlier Snowflake categories[4] (Snowflake, Snowflake plus and minus) to include two new continuum categories called the "Asymmetric Snowflake plus and minus". We will call this theoretical framework MagVAC and will use it as a reference.

In this paper, we develop a totally different approach based on an investigation and examination of the detailed structure of the magnetic field of advanced divertors in only the physically relevant region for power exhaust – the Scrape-Off Layer (SOL) terminating on the divertor plates. After all, what is relevant for divertor physics, and for the divertor's ability to distribute the heat load, is the structure of the magnetic field in the SOL where heat is conducted. The analysis presented in this work, concentrating on the magnetic field only in the heat exhausting SOL, will be called MagSOL; it employs qualitative physical arguments, extensive numerical work, and a detailed analysis using the model magnetic field of Ref. 6.

As hoped, MagSOL analysis leads to the emergence of a physical "metric", the Divertor Index $DI$, that allows us to distinguish between the three geometries of interest: the Standard Divertor (SD, $DI = 1$), and the two advanced geometries – the X-Divertor (XD, $DI > 1$) and the Snowflake (SFD, $DI < 1$). The MagSOL approach based on the shape of SOL flux surfaces is similar in spirit to the classification of core plasma shapes into circular, elongated, and oblate – independent of fields and coils outside the plasma.

Within the context of the MagSOL framework, we find that experimentally successful geometries on NSTX and DIII-D have the characteristics and properties of the 2004 XD; they have been simply mislabeled as Snowflakes. Such errors in labeling the configurations can lead to errors in judgments on their projected physical properties for future development of advanced divertors.

MagSOL, we believe, may prove to be vital in providing a theoretical basis for advanced divertors. Such a detailed analytical, geometrical, and physical examination of advanced divertors is expected to lead to: 1) physics-based quantitative metrics, like $DI$, that encapsulate the intuitive ideas and make it easy to identify different advanced divertors, 2) the clarification of characteristics that each configuration is (and is not) expected to have, and 3) the exposure of many subtle issues that may have led to the mislabeling of recent experiments.

## II. Overview

Because of the complexity of the subject, we have chosen to give an overview of the paper before giving a detailed treatment of various issues in Sections III-VI.

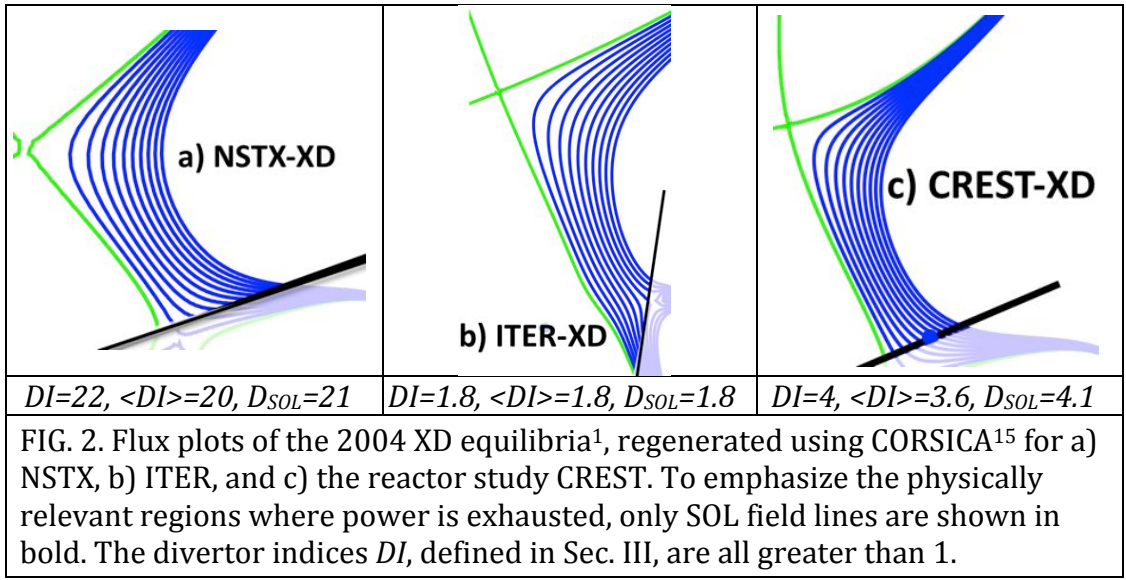

| $DI=22, <DI>=20, D_{SOL}=21$ | $DI=1.8, <DI>=1.8, D_{SOL}=1.8$ | $DI=4, <DI>=3.6, D_{SOL}=4.1$ |

FIG. 2. Flux plots of the 2004 XD equilibria[1], regenerated using CORSICA[15] for a) NSTX, b) ITER, and c) the reactor study CREST. To emphasize the physically relevant regions where power is exhausted, only SOL field lines are shown in bold. The divertor indices *DI*, defined in Sec. III, are all greater than 1.

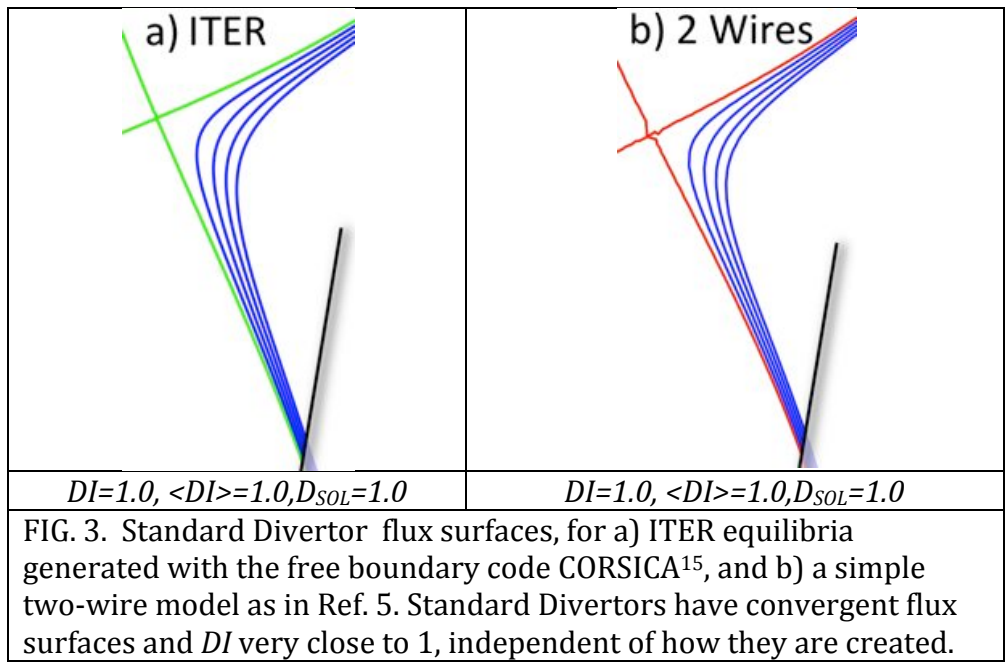

| $DI=1.0, <DI>=1.0, D_{SOL}=1.0$ | $DI=1.0, <DI>=1.0, D_{SOL}=1.0$ |

FIG. 3. Standard Divertor flux surfaces, for a) ITER equilibria generated with the free boundary code CORSICA[15], and b) a simple two-wire model as in Ref. 5. Standard Divertors have convergent flux surfaces and *DI* very close to 1, independent of how they are created.

### A. A brief history of AD research

The history of AD research is marked by three distinct developments:

1) The X-Divertor (XD), was introduced in 2004[1]. In Fig.2, we display a few of the representative 2004 XD configurations from the original paper; a) NSTX, b) ITER, c) CREST. The defining and most important characteristics of the XD were stated in the abstract of the introductory paper[1], and in the text therein:

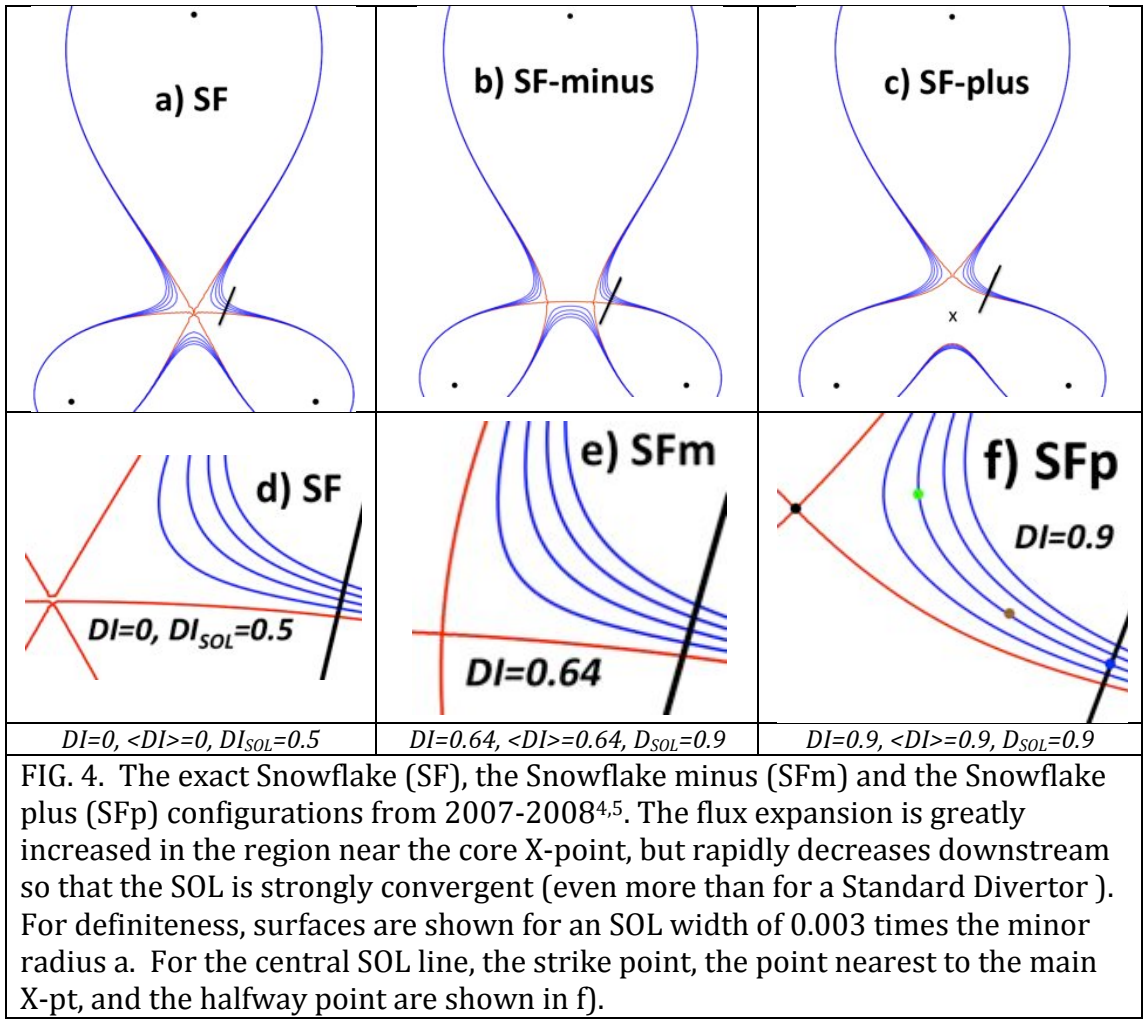

| $DI=0$, $\langle DI\rangle=0$, $DI_{SOL}=0.5$ | $DI=0.64$, $\langle DI\rangle=0.64$, $D_{SOL}=0.9$ | $DI=0.9$, $\langle DI\rangle=0.9$, $D_{SOL}=0.9$ |
|---|---|---|

FIG. 4. The exact Snowflake (SF), the Snowflake minus (SFm) and the Snowflake plus (SFp) configurations from 2007-2008[4,5]. The flux expansion is greatly increased in the region near the core X-point, but rapidly decreases downstream so that the SOL is strongly convergent (even more than for a Standard Divertor ). For definiteness, surfaces are shown for an SOL width of 0.003 times the minor radius a. For the central SOL line, the strike point, the point nearest to the main X-pt, and the halfway point are shown in f).

a) The XD configuration is created by "inducing a second axisymmetric X-point downstream of the main plasma X-point"[1].
b) The beneficial result that "field line lengths from the core X-point to the wall can be increased… and flux expansion can be increased". The physical consequence of these characteristics was predicted to be a greatly reduced heat flux on the divertor plate. It was also suggested in Ref. 1 that the XD may allow a stable detached operation.
c) The new X-point produced a new geometry in the SOL flux surfaces – the flux surfaces flared outward, rather than contract inward as in a Standard Divertor . (See Fig.3 for typical SD examples).
d) The name X-Divertor succinctly describes the physical essence of the configuration – the downstream SOL interacting with a new X-point.

2) In subsequent years (2007 onwards), a few other advanced divertor configurations have been proposed. The most relevant ones are:

a) The Snowflake Divertor (SFD) family, built around the basic configuration (to be called pure Snowflake) that creates a *second order null* at the main plasma X-point, was introduced in 2007. Such a second order null in the core X-point implies a six-fold symmetry in the magnetic field in the divertor region, again, leading to the apt and succinctly descriptive name "Snowflake". The pure Snowflake was complemented by two variants – "plus" and "minus" in the first publications on the subject[4,5] (See Fig.4). These variants do not have one second order null; instead they have two first order nulls (like the XD), but placed relative to each other so that the characteristics of the pure Snowflake (for example, enormous flux expansion and increase in the line length in the vicinity of the two nearby X-points) may be maintained while its major problem – instability to small perturbations – is overcome. In the rest of the paper, the term SFD will be used to denote only these variants. The original Snowflake papers were quite explicit[4,5] in stating that SFDs were different from XD and pointing out the differences[5]. Note, for example, that the flux surfaces in the power-exhausting SOL region are all convergent in Fig.4 – the opposite of the 2004 X-Divertor examples displayed in Fig.2.

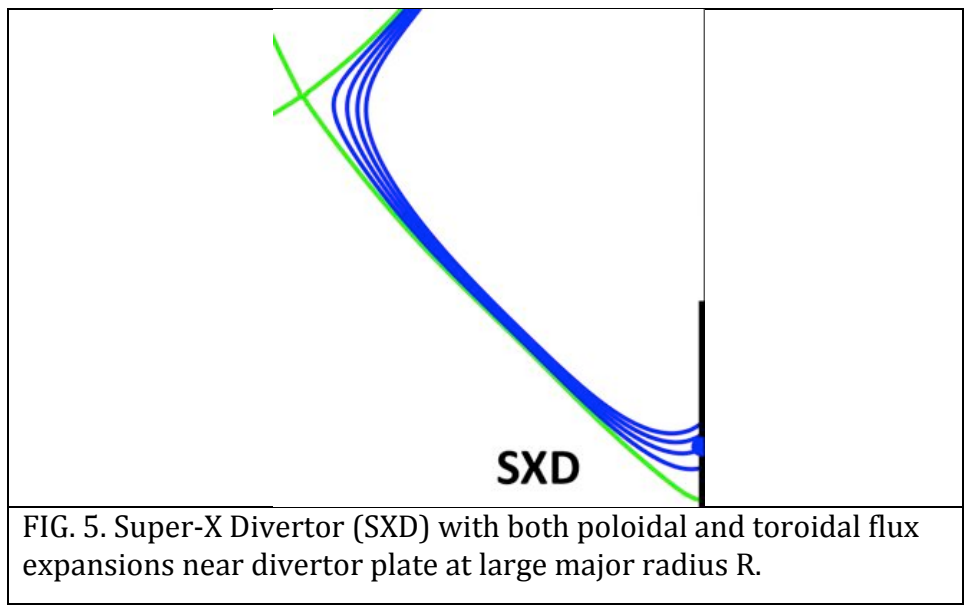

FIG. 5. Super-X Divertor (SXD) with both poloidal and toroidal flux expansions near divertor plate at large major radius R.

b) In the same year as the Snowflake (2007), a more advanced variant of the XD, named the "Super-X" divertor (SXD, Fig.5), was presented; the latter attains even a greater flux expansion by superposing toroidal expansion (by placing the divertor plate at the maximum possible major radius) on the poloidal flux expansion[7,8,9]. Once again, the choice of the name, quite accurately, reflects the intended function: modification of the X-Divertor concept for "superior" flux expansion and line length. In this paper, however, we will not discuss the SXD but concentrate on the XD and the Snowflake variants that rely only on poloidal flux expansion. In the rest of this paper the notation XD will exclude SXD, and will refer only to the class of configurations (this will be elaborated later) that correspond to the AD of 2004[1].

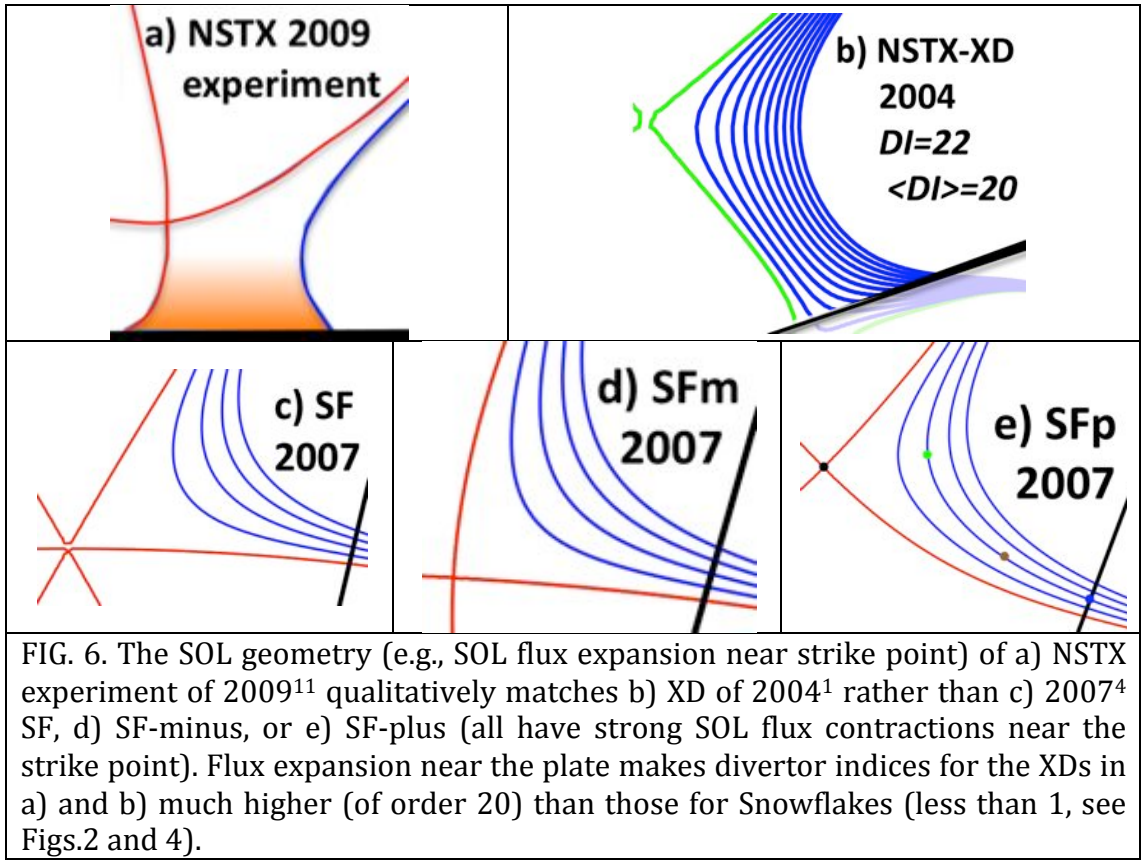

FIG. 6. The SOL geometry (e.g., SOL flux expansion near strike point) of a) NSTX experiment of 2009[11] qualitatively matches b) XD of 2004[1] rather than c) 2007[4] SF, d) SF-minus, or e) SF-plus (all have strong SOL flux contractions near the strike point). Flux expansion near the plate makes divertor indices for the XDs in a) and b) much higher (of order 20) than those for Snowflakes (less than 1, see Figs.2 and 4).

3) These conceptual developments of advanced divertors stimulated a robust experimental program to create the configurations on the current tokamaks. *The Snowflake authors played an essential role in motivating the highly successful experiments carried out at TCV, NSTX, and DIIID in 2009-2012* (Figs. 1 and 6 show these geometries). Although we will present an analytical framework to interpret the recent experiments, here is a very brief summary of the relevant published experimental results that were all labeled Snowflakes:

a) The TCV experiments had SOL flux surfaces that contract away from the core plasma X-point, whereas successful configurations on NSTX and DIII-D had flared flux surfaces near the strike points.
b) The TCV experiments reported enhanced radiation near the main X-point[10], while the NSTX/DIIID found the radiation to be localized near the divertor plate[12,13].
c) All three machines reported good flux expansion and increased line length, and appropriately reduced heat fluxes on their divertor plates.

4) After the successful NSTX experiments in 2009, two new "extended" Snowflake categories were introduced in Ref. 6 – the "Asymmetric Snowflake minus" and "Asymmetric Snowflake plus". The experimentalists have labeled NSTX/DIIID configurations as "Asymmetric Snowflake minus".

We show in this paper that these advanced geometries are X-Divertors (see Fig.6), and should be named as such.

**B. Outline of the Paper**

The most important goal of this paper is to develop what we will call MagSOL, a framework based on SOL physics, similar to the one that is used to categorize the core plasma geometry, to better analyze divertor geometry and physics.

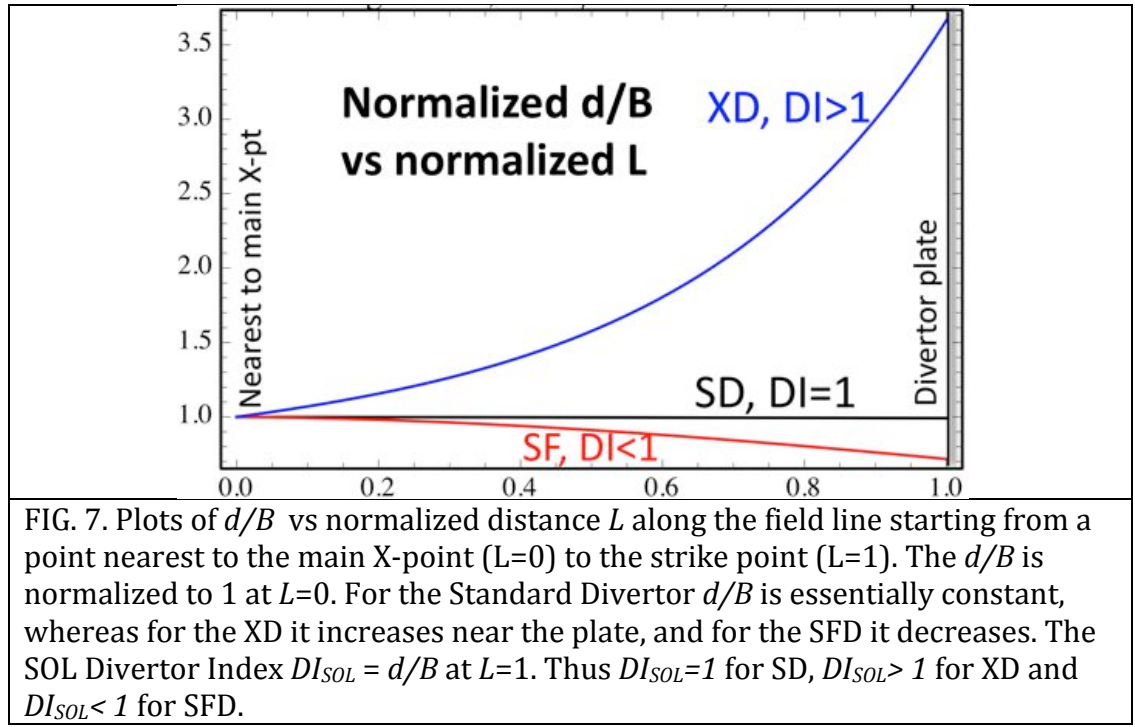

FIG. 7. Plots of $d/B$ vs normalized distance $L$ along the field line starting from a point nearest to the main X-point (L=0) to the strike point (L=1). The $d/B$ is normalized to 1 at $L$=0. For the Standard Divertor $d/B$ is essentially constant, whereas for the XD it increases near the plate, and for the SFD it decreases. The SOL Divertor Index $DI_{SOL} = d/B$ at $L$=1. Thus $DI_{SOL}$=1 for SD, $DI_{SOL}> 1$ for XD and $DI_{SOL}< 1$ for SFD.

In Sec. III, we will begin with a critical examination of the structure of the magnetic field in the SOL. The scrutiny will involve qualitative physics arguments, extensive numerical work, and a detailed analysis using the model magnetic field given in Ref. 6. We will introduce the Divertor Index *DI*, a new "metric" signifying the increase of flux expansions as one goes from the main X-point to the strike point (Fig.7).
The index *DI* has another simple physical interpretation: near the strike point, the SFD flux surfaces converge more rapidly than the SD while for the XD they diverge more rapidly than the SD.

According to these criteria, the successful configurations given in NSTX/DIIID publications look and behave precisely like what was predicted for the XD (compare Fig.1a-c with Fig.2), whereas the TCV experiments do, indeed, belong to the SFD family of configurations (compare Fig.1d-f with Fig.4c-e). We will show that the predictions of the MagSOL analysis are in contradiction to the categorization and nomenclature used by some experimental groups.

In Sec. IV, we will discuss the implications of MagSOL for divertor physics, in particular, the physics and dynamics of detachment. We will show that the XD magnetic geometry is particularly suited to explore the possibility of a stable detached operation as speculated and suggested in Ref. 1.

In Sec. V, we will discuss the very important issue of the significance and the extrapolation potential of the current experiments to ITER and future reactors. Extrapolation assumes an extra degree of difficulty since the SOL power width could change by an order of magnitude as we progress, say, from NSTX to ITER. We will show that for a dependable extrapolation of divertor action, one must insist on "replicating" the magnetic field structure in the SOL; similar SOLs yield similar physics. Thus the characteristic SOL-based metrics, introduced in this study, will provide powerful tools for insuring beneficial divertor action in future machines.

Section VI will conclude with a summary.

## III. Analyzing Divertor Geometries – Distinguishing Metrics

### A. General Framework: The plasma SOL as the relevant region for divertor action

It is worth stressing at this point that the physically relevant divertor region, delimited by both the SOL width and the position of the divertor plate, typically occupies quite a *small* fraction of the total region that encompasses the divertor, the second X-point, the divertor coil region, etc. However, a physically meaningful classification scheme must focus on this region; only then such a scheme could aspire to connect physical behavior with divertor category, and to allow a reliable extrapolation of divertor behavior to different devices with very different $\Delta/a$, where $\Delta$ is the SOL width. Recall that $\Delta/a$ on ITER and reactors is expected to be many times less than on the present experiments[16,17].

We must, therefore, specify the SOL width in terms of the relevant flux surfaces to exhaust power from the main plasma. Specification of the SOL width is, of course, an additional complication; and it is a source of uncertainty when extrapolating. This complication has to be necessarily dealt with because the divertor performance is a consequence of the interaction of the SOL plasma with local magnetic geometry in the SOL; it cannot be uniquely extrapolated based on the magnetic field structure in regions away from the SOL.

When sufficient experimental data at the divertor plate is available, the SOL width can be determined after the fact: by mapping the divertor heat flux and/or divertor radiation upstream along field lines. Alternatively, if acceptably accurate formulas are available for the upstream SOL width, it could be mapped downstream to define the SOL. Recent formulas by Goldston[16,17] are promising in this respect.

Finally, the position of the divertor plate must be specified. It defines the end of the SOL. The region beyond the plate is outside the plasma and any field in that region has no influence on divertor action. For projections to future devices, this also entails specifying more than the magnetic field alone.

**B. Conceptual framework for MagSOL: The Divertor Index DI – Relative Flux Expansion and Relative Convergence (or Divergence) Rates**

Let us consider the field structure implied by the magnetic field geometries labeled X-Divertor and Snowflake. The X-Divertor introduces a second X-point "downstream"[1]. Evidently, this *preferentially modifies the magnetic fields in the region downstream* – i.e., away from the main plasma X-point. If we take this concept to its limit, the region most affected will be the region *most* downstream where the SOL contacts the divertor plate – the SOL strike point. This is in contrast to the Snowflake strategy in which a second order null is introduced at the core plasma X-point: the six-fold symmetry that motivates the name "Snowflake" is the direct consequence of this. It is emphasized in Refs. 4-6 that the Snowflake configurations modify the Standard Divertor , *preferentially near the core X-point*.

We have used the term 'preferentially' because it is inevitably true that introducing a new X-point modifies the magnetic field, to some extent, in both regions. Hence, the X-Divertor configurations (Fig.2), with the X-point near the plate, also reduce the magnetic field around the core plasma X-point (as was recognized by the authors in 2004). Likewise, the Snowflake configurations (Fig.4), with a second order null (exact or approximate), reduce the magnetic field downstream at the plate, for appropriate plate positions. Nonetheless, since the prescriptions for the XD and SFD emphasize different regions of the SOL, it is relatively straightforward to construct physically based metrics that distinguish them.

One quantity of obvious physical interest is the flux expansion – essentially the reciprocal of the poloidal magnetic field. We can distinguish the XD and Snowflake by whether the flux expansion is modified from a Standard Divertor preferentially near the core X-point, or mainly downstream (e.g. at the plate). This distinction leads directly to the visually obvious difference between the XD and the SFD: whether the field lines (near the plate) are less or more convergent than SD.

With this as background, we can now quantify the visual difference that would stem from the differing prescriptions for XD and SFD. Consider two positions *a* and *b*, where b is the downstream terminus of an SOL field line on the divertor plate, and *a* is the position on that same field line that is closest to the core X-point (see Fig.8). The ratio of the flux expansion at *b* to its value at *a* is $B_a/B_b$. (Throughout this paper we will use the symbol *B* to denote the poloidal magnetic field). Both the Standard Divertor and the pure Snowflake have converging flux surfaces: they differ, however, in the rate of convergence. The Standard Divertor magnetic fields $B_{SD}$ varies *linearly* with distance *d* from the core X-point. Hence, convergence of flux surfaces *relative to a Standard Divertor* , is given by the SOL Divertor Index ($DI_{SOL}$):

$$DI_{SOL} \equiv \frac{d_b/B_b}{d_a/B_a} = \frac{B_a}{B_b}\frac{d_b}{d_a} \tag{1}$$

If $DI_{SOL} > 1$, the flux surfaces are more flared than a Standard Divertor , and if $DI_{SOL}<1$, it is more contracting than a Standard Divertor . Thus, $DI_{SOL}$ is the numerical quantification of the visual criterion stated earlier.

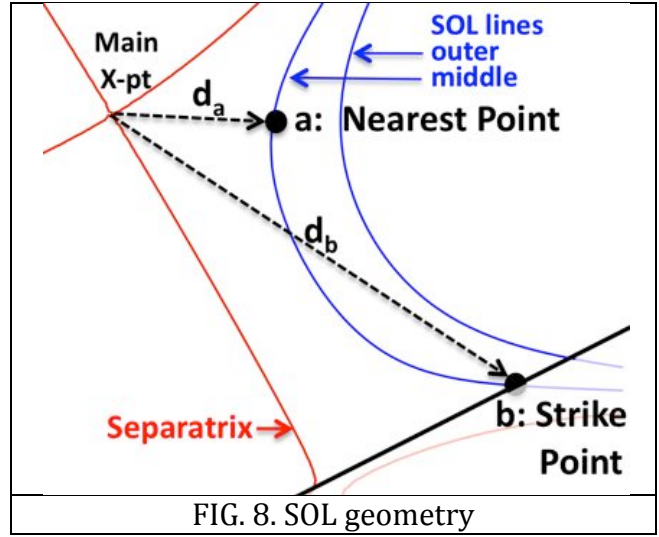

FIG. 8. SOL geometry

Note that the cross sectional area of the SOL is proportional to R/B. The R variation is small for the divertors considered here (since we are not discussing the SXD), so for clarity we will neglect it in what follows.

One must select a specific field line to evaluate $DI_{SOL}$. For simplicity, to select a single field line representative of the whole SOL, we can take a line in the "middle" of the SOL. For example, suppose the SOL width is specified as in the formulas by Goldston[16] as some value $\Delta$ from the separatrix on the outboard mid-plane. Then, for the middle of the SOL, take a field line mapped downstream from the mid-plane, starting a distance $\Delta/2$ from the outboard separatrix.

Alternatively, one can consider the entire bundle of field lines starting within the width $\Delta$ from the outboard mid-plane, and map them to the divertor plate, which defines the "wetted area". One could then compute the integral average of $DI_{SOL}$ over the length of the wetted area, $<DI_{SOL}>$.

An alternative and illuminating interpretation of $DI_{SOL}$ is revealed through a "thought experiment". We start with a Standard Divertor , and modify the magnetic fields to create an advanced divertor. The SD magnetic field $B_{SD}$ varies *linearly* with distance $d$ away from the core X-point, so for any two positions labeled by a and b

$$B_{SDb} \approx B_{SDa}(d_b / d_a) \tag{2}$$

Now, we modify the Standard Divertor, by changing the magnetic field, to create an advanced geometry. Consider some field line in the relevant SOL. Consider the position along this field line that is closest to the core X-point. The SOL magnetic field at this position is now $B_a$. The local flux expansion has been increased, relative to the Standard Divertor, by the ratio $F_a = B_{SDa}/B_a$. Downstream, at position b (e.g. the terminal point on the plate), the flux expansion has been increased, again relative to a Standard Divertor, by the ratio $F_b = B_{SDb}/B_b$. We can now ask the question: is the flux expansion increased primarily near the plate, or nearest to the core X-point?

Such physically pertinent questions lead to exactly the same index as in Eq. 1

$$(F_b / F_a) = (B_a / B_b)(d_b / d_a) = DI_{SOL} \tag{3}$$

An XD primarily increases flux expansion downstream at position *b*: $DI_{SOL} > 1$. A Snowflake preferentially increases flux expansion at position *a*: $DI_{SOL} < 1$.

An alternative quantity *DI*, which is somewhat simpler to compute in practice, can also be defined. As we take the limit where *a* approaches the core X-point, $B_a/d_a \rightarrow |\nabla B|$ at the core X-point. We can then define

$$DI \equiv |\nabla B|(d_b / B_b) \tag{4}$$

It can be shown that $|\nabla B|$ is just the square root of the Jacobian at the core X-point,

$$|\nabla B| = |(dB_R/dR)(dB_Z/dZ) - (dB_R/dZ)(dB_Z/dR)|^{1/2} \tag{5}$$

where *R* and *Z* are the usual cylindrical coordinates that are employed for axi-symmetric tokamak geometries.

*For a pure Snowflake, DI = 0. For a Standard Divertor, DI = 1. For a "pure" XD where the second X-point is located at the strike point, DI approaches infinity.*

It follows, then: 1) If *DI > 1*, the configuration is primarily an XD, and if *DI < 1*, the configuration is primarily an SFD, and 2) The XD has the maximum flux expansion at the plate while the maximum flux expansion for the SFD is at the main X-point.

Notice that both *DI* and *DI$_{SOL}$* depend on the core X-point and the plate – the two regions emphasized in the prescriptions for an XD and an SFD. Both *DI* and *DI$_{SOL}$* are computed solely from values of the magnetic field in the power exhausting SOL. They do not use field properties in any irrelevant region outside the plasma domain.

Because there are several ways to evaluate the Divertor Index *DI*, we have chosen to display three versions – *DI*, *<DI>, DI$_{SOL}$*, on the figures; we find perfect correlation between these quantities and the geometries from the respective XD and SFD papers. Note that *<DI$_{SOL}$>* is nearly always quite close to *<DI>*, so sometimes only the latter is displayed. As can also be seen, these values correspond to the visually apparent (more convergent or flared than SD) shapes of the SOL.

Recall the analogy with the core plasma geometry, which is specified by the shape. The plasma elongation can be quantitatively specified either as 1 (a circular plasma, the reference case), > 1 (an elongated plasma), or < 1 (an oblate plasma). The quantities *DI* are analogous to elongation for the core plasma.

The level of convergence or divergence of SOL exhaust flux surfaces has important effects on the operational behavior (and the expected physical behavior), and as noted above, these differences of the SOL magnetic field in the exhaust region can be distilled into specific, well defined quantities. However, let us delve a little more into the geometry of advanced configurations by considering plots of the flux expansion along the entire length of the exhaust SOL field lines.

The Snowflake and the Standard Divertor both have convergent flux surfaces, but differ only in the rate of convergence. It is difficult to distinguish them by visual inspection using a plot of *1/B* alone. For visual clarity, one can plot a quantity that makes the *rate* of convergence more readily apparent. So we plot *d/B* , where d is the distance from the main plasma X-point. This choice allows easy visual differentiation of all geometries (See Fig.7):

1) For a Standard Divertor , *1/B~ 1/d*. If *d/B* turns out to be approximately constant, we have a Standard Divertor .

2) The Snowflake is constructed by producing a second order null, either exactly or approximately. For a higher order null, the flux surfaces converge more rapidly than for a Standard Divertor . Thus if d/B decreases with distance for a path along an SOL field line, we have a Snowflake variant.

3) Conversely, if, relative to a Standard Divertor , *d/B* is an increasing function of distance, then the underlying geometry conforms to the X-Divertor configuration proposed and discussed in 2004.

4) The quantity *DI$_{SOL}$* is the ratio of the d/B at the end points of the graph above. These *DI* distill the gross geometrical behavior of the plot into a single number.

The *d/B* versus *L* plots along the line in Fig.7, encompassing all divertor configurations under scrutiny (Figs.1-4), convey the principal message of this section: there is a clearly visible qualitative splitting between the three divertors (Fig.7): The Standard Divertor s follow an essentially constant line, the X-Divertors follow a curve that is clearly increasing as the magnetic field line travels toward the divertor plate, and the Snowflakes ride a curve that is decreasing as the field line moves toward the plate. The XD and the 2007 SFD lie on opposite sides of the Standard Divertor , and represent modifications of the Standard Divertor in diametrically opposite directions.

**C. Analysis of a Model Magnetic Field – MagVAC and MagSOL**

The earlier divertor classification scheme, MagVAC of Ref. 6, is anchored in the study of the magnetic field in the divertor region using a Taylor expansion of the vacuum field near the main X-point. The analytic representation is accurate enough in the region of interest as long as the coils are relatively far away. All novel divertor configurations with two X-points sufficiently close to each other (the main X-point is located at $x = z = 0$, while the second X-point is at ($X_2, Z_2$) at a distance $d_{xpt} = (X_2^2 + Z_2^2)^{1/2}$ away from the main X-point) are analyzed by exploring the generic approximate vacuum (poloidal) $B$ field

$$\vec{B} \approx \vec{e}_x[x^2 - z^2 + (zZ_2 - xX_2)] + \vec{e}_z[-2zx + (xZ_2 + zX_2)] \tag{6}$$

of which

$$\vec{B}_{SF} \approx \vec{e}_x[x^2 - z^2] - 2zx\vec{e}_z \tag{7}$$

is the field of an ideal Snowflake with a second order null.

The poloidal magnetic field can be expressed in polar coordinates defined as $x = d Cos\theta$, $z = -d Sin\theta$, and $X_2 = d_{xpt} Cos\theta_{xpt}$, $Z_2 = -d_{xpt} Sin\theta_{xpt}$. Due to reflection symmetry, the entire range of possible X-point angles can be taken to be within $0 < \theta_{xpt} < \pi/2$.

We begin with a brief summary of the classification scheme MagVAC :

1) Far away compared to the distance between two X-points ($d >> d_{xpt}$), the flux surfaces have a six-fold character like the original Snowflake configuration[4]. This similarity is, apparently, the only shared property that supports applying the name Snowflake (with appropriate additional qualifiers) to all advanced configurations with two X-points, i.e., to all configurations accessible to the field described by Eq. 6.

2) Two new qualified names (beyond the 2007 variants, the "Snowflake plus" [$\theta_{xpt} = \pi/2$] and "Snowflake minus" [$\theta_{xpt} = 0$]) of the Snowflake are introduced: when the second X-point ($d_{xpt}, \theta_{xpt}$) lies in the private region, the configuration is described as an "Asymmetric Snowflake plus", otherwise, the configuration is an "Asymmetric Snowflake minus". With this generalization, the entire range $0 < \theta_2 < \pi/2$ would be covered by the extended Snowflake category.

Since the X-Divertor "introduces a second axi-symmetric X point downstream" the 2004 XD[1] becomes, *inevitably*, for some angle $\theta_{xpt}$, one of the members of the extended Snowflake family. In particular, the 2004 XD configurations of Fig.2 (the second X-point is somewhat close to the main plasma separatrix, but not in the private region) would, then, be an example of an "Asymmetric Snowflake minus".

In the analysis in Ref 6, the physically relevant region in the exhaust SOL is not specifically examined in detail. Because of this, there is a disconnect between physically relevant quantities and the mathematical development in Ref 6. We will now present a distinctly different mathematical development that will be closely connected to physical variables of interest to divertor action.

Let us examine the magnetic field of Eq. 6 to find answers to two basic questions:

1. How much of *B* is "Snowflake-like" – has the properties of the pure Snowflake? This will be a simple magnitude measure.
2. Does *B*, for any position of the second X-point, represent a unique or a universal geometry so that a single name (Snowflake, for instance) could be useful, or does it encompass very different geometries that might demand a richer classification?

Since we are studying the physical problem of divertor performance, both the geometry as well as magnitudes must and will be examined only in the region occupied by the SOL plasma.

To answer the first question, we define, what may be called a "Departure Function"

$$DF = \frac{\left|\vec{B} - \vec{B}_{SF}\right|}{\left|\vec{B}_{SF}\right|} \tag{8a}$$

which measures the fractional magnitude of the non-Snowflake-like component over the Snowflake-like (SF) component. The Departure Function DF, calculated exactly for the model magnetic field of Eq. 6 is simple but particularly revealing:

$$DF = \sqrt{\frac{X_2^2 + Z_2^2}{x^2 + z^2}} = \frac{d_{xpt}}{d} \tag{8b}$$

- The *DF* depends only on the distances ($d, d_{xpt}$) from the main X-point – independent of the angular location of the second X-point.
- For any non-zero $d_{xpt}$ the Departure Function *DF > 1* for all $d < d_{xpt}$ and *DF* becomes very large at $d \ll d_{xpt}$. Thus in the vicinity of the main X-point, i.e., at $d < d_{xpt}$, the field is always predominantly unlike the field of a pure Snowflake.

This region shows a *discontinuous* and strong departure from the singular case of a pure Snowflake with $d_{xpt} = 0$; we remind the reader that Ref. 6 uses this very singular field to name the entire class.

- In the experiments on NSTX and DIIID, and in all the published XD cases, the physically relevant SOL region (that ends at the divertor plate) is defined by $d_{max} \leq d_{xpt}$. Thus, *in the region of physical interest*, the $\vec{B}$ (Eq. 6) analyzed in Ref. 6 contains too little "Snowflake-component" to warrant "Snowflake-like" label.
- If, however, $d_{max}$ is considerably larger than $d_{xpt}$, the field is predominantly Snowflake-like and the configuration should be certainly called Snowflake. This situation pertains, for instance, in reported TCV experiments where $d_{max} >> d_{xpt}$.

Note that for the typical position of an X-point in an XD and in the experimental cases on NSTX and DIII-D, the region where the Snowflake terms dominate is mostly outside the plasma – beyond the divertor plate, i.e., not a region of physical interest.

The preceding exercise may drive us to the conclusion that, barring the topologically unstable pure Snowflake (one second order null), none of the configuration with two separate single-order X-points contained in the simple analytical formula Eq. 6 could be legitimately called Snowflakes except at distances $d_{max} >> d_{xpt}$. This, however, would not be correct; there is, indeed, a non-negligible Snowflake component in the domain of interest that can lead to a qualitatively Snowflake-like structure for some parameters, even when $d_{max} \sim d_{xpt}$.

We now attempt a more detailed analysis to find the appropriate region (amongst all the physically relevant configurations) where the name "Snowflake-like" is appropriate. This detailed analysis should also help us to identify the regime of other two X-point geometries like the X-Divertor.

This detailed examination of the analytic magnetic field was inspired by the more conceptual/geometric arguments presented in Sec. IIIb; the analysis, in turn, provides, *a fortiori*, mathematical basis for the intuitive elements of MagSOL.

At this stage, we bring the SOL into the analysis by introducing the strike point ($d_s, \theta_s$). Using the formula for the total magnetic field of Eq. 6 in polar coordinates,

$$\frac{B^2}{d^2 d_{xpt}^2} = 1 + \frac{d^2}{d_{xpt}^2} - 2\frac{d}{d_{xpt}}Cos(\theta - \theta_{xpt}) \tag{9}$$

we can derive an analytic expression for the Divertor Index *DI* for the model field:

$$DI = \frac{(d/B)_{(d_s, \theta_s)}}{(d/B)_{(0.0)}} = \left[1 + \frac{d_s^2}{d_{xpt}^2} - 2\frac{d_s}{d_{xpt}}Cos(\theta_s - \theta_{xpt})\right]^{-1/2} \tag{10}$$

**Snowflakes: *DI* < 1,  Standard Divertor : *DI* = 1, X-Divertors: *DI* > 1**

| Divertor | Definition | Metric | Comments |
|---|---|---|---|
| X-Divertor (XD) | $(d_s, \theta_s) \approx (d_{xpt}, \theta_{xpt})$ <br> $d_s / d_{xpt} < 2 Cos(\theta_s - \theta_{xpt})$ | $DI_{XD} \approx \left\|(1 - d_s / d_{xpt})\right\|^{-1} > 1$ | For all $0 < d_s / d_{xpt} < 2$. Ideal XD has $d_s \rightarrow d_{xpt}$ so that $DI_{XDI} \rightarrow \infty$ |
| Standard | $d_{xpt} \rightarrow \infty$ | $DI_{SD} \approx 1$ | for all $\theta_s$ and finite $d_s$ |
| Snowflake (SF) | $d_s >> d_{xpt}$ <br> $d_s / d_{xpt} > 2 Cos(\theta_s - \theta_{xpt})$ | $DI_{SF} \approx d_{xpt} / d_s < 1$ | for all $\theta_s$, approaches ideal (singular) Snowflake when $d_{xpt} \rightarrow 0$. |

TABLE I. Comparison of Standard, X, and Snowflake Divertors.

This follows, exactly, the more intuitive arguments of Sec. IIIb. Table I compares the various definitions and metrics for the three classes of divertors.

Formula 10, approximate as it is for a real configuration, serves as a very interesting tool for understanding the structure of magnetic geometry. The Divertor Index *DI* is determined by $d_s / d_{xpt}$ and $\theta_s - \theta_{xpt}$, both are hybrid quantities born out of the interplay of the magnetic geometry with the SOL. Notice that $\theta_{xpt}$, the angular location of the second X-point has no relevance by itself; it is only the relative angle $\Delta\theta = \left|\theta_s - \theta_{xpt}\right|$ between the X-point and strike point location that matters.

We end the analytic examination by displaying the very simple expression for the flux expansion $F_{d_s = d_{xpt}} \sim 1/B$ at $d_s / d_{xpt} = 1$,

$$F_{d_s = d_{xpt}} \sim \frac{1}{2 d_{xpt}^2 Sin(\left|\Delta\theta\right|/2)} \quad (11)$$

which is boosted to large values for small $\Delta\theta$, the domain claimed by the 2004 XD. Notice that, in this range, flux expansion can be strongly increased by bringing the strike area close to the second X-point.

Hence, modest changes in the position of the second X-point relative to the strike point can have a stronger effect on the flux expansion at the plate than modest changes in the *d*<sub>xpt</sub>. Therefore, there is no basis to consider small *d*<sub>xpt</sub> as crucial for producing an advanced divertor. The position of X-point near the plate, as indicated in the XD papers, has a much larger quantitative effect on flux expansion.

Also notice that $DI_{d_s = d_{xpt}}$ changes its "nature" – it goes from > 1 to < 1 at

$$sin(|\Delta\theta|/2) = 1/2 \quad (12)$$

For $d_s \sim d_{xpt}$, if the angular separation $\Delta\theta$ is less than $\pi/3$, the geometry is that of an X-Divertor, and if $\Delta\theta$ is more than $\pi/3$, it is a Snowflake. The changeover value of $\Delta\theta$ will, of course, change if $d_s$ and $d_{xpt}$ are different. For each choice, there will be well-defined and separated domains for the XD and the Snowflake geometries.

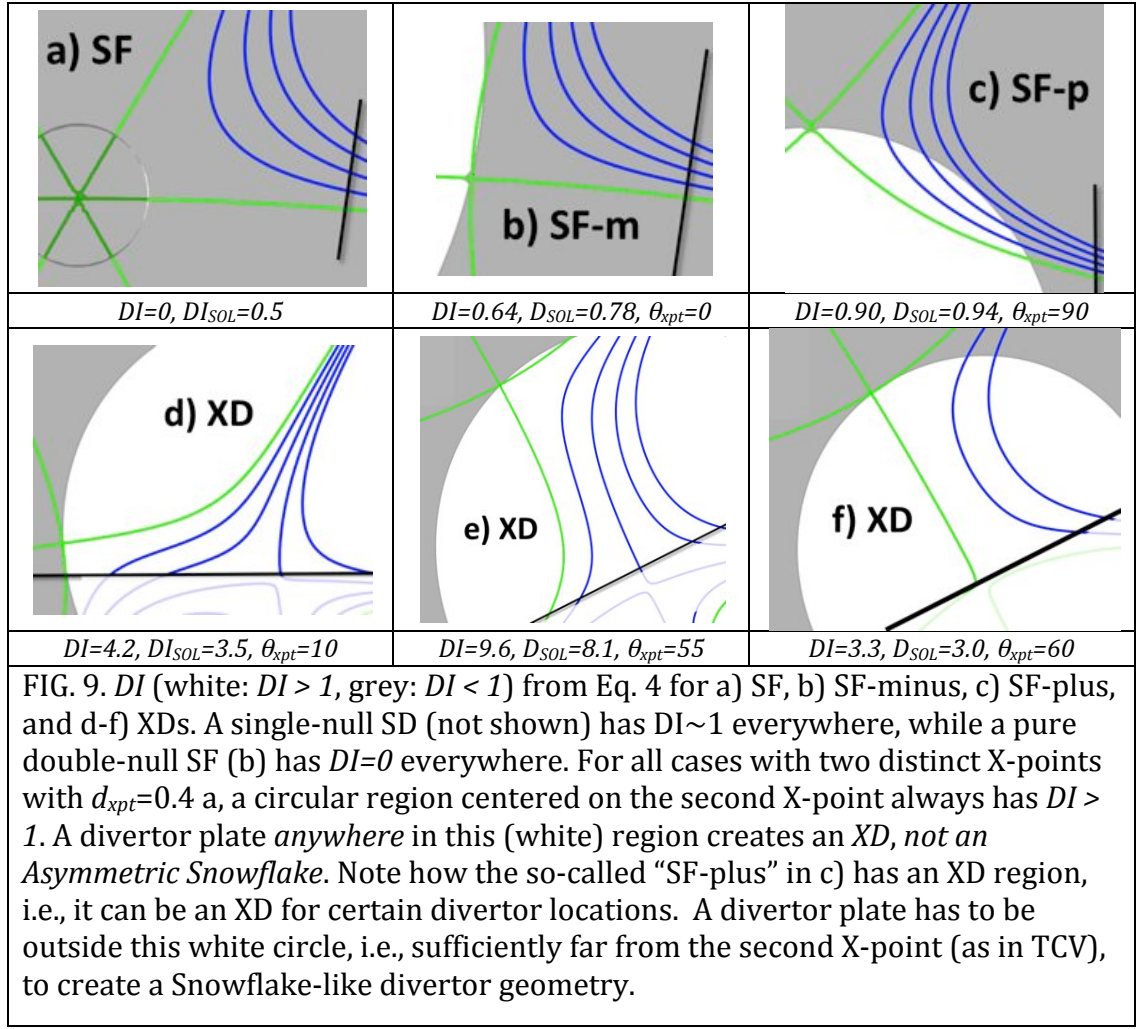

| | | |
|---|---|---|
| *DI=0, DI$_{SOL}$=0.5* | *DI=0.64, D$_{SOL}$=0.78, θ$_{xpt}$=0* | *DI=0.90, D$_{SOL}$=0.94, θ$_{xpt}$=90* |
| *DI=4.2, DI$_{SOL}$=3.5, θ$_{xpt}$=10* | *DI=9.6, D$_{SOL}$=8.1, θ$_{xpt}$=55* | *DI=3.3, D$_{SOL}$=3.0, θ$_{xpt}$=60* |

FIG. 9. *DI* (white: *DI > 1*, grey: *DI < 1*) from Eq. 4 for a) SF, b) SF-minus, c) SF-plus, and d-f) XDs. A single-null SD (not shown) has DI~1 everywhere, while a pure double-null SF (b) has *DI=0* everywhere. For all cases with two distinct X-points with $d_{xpt}$=0.4 a, a circular region centered on the second X-point always has *DI > 1*. A divertor plate *anywhere* in this (white) region creates an *XD*, *not an Asymmetric Snowflake*. Note how the so-called "SF-plus" in c) has an XD region, i.e., it can be an XD for certain divertor locations. A divertor plate has to be outside this white circle, i.e., sufficiently far from the second X-point (as in TCV), to create a Snowflake-like divertor geometry.

Our analytical results and (the conceptual antecedents) are verified by numerical calculations of divertor fields using the model two wire coils as invoked in the Snowflake papers. The plots displayed in Fig.9b-f consider geometries where $d_{xpt}$ is about 0.4 times the minor radius a – about the same as in 1) experiments (where the Ref. 6 characterization is applied), 2) in the 2004 XD paper, 3) the 2007 SFD examples of "plus" and "minus", and 4) the distance on ITER from the main plasma X-point to the strike point on the divertor plate.

We find, as expected:

1) For $\theta_{xpt}$ close to 90°, and for appropriate values of $d_s/d_{xpt}$, and $\Delta\theta = |\theta_s - \theta_{xpt}|$, the configuration is clearly similar to a Snowflake plus. Hence the designation adopted in Ref. 6 -"Asymmetric Snowflake plus" – is appropriate.
2) For $\theta_{xpt}$ close to 0°, and commensurate $d_s/d_{xpt}$, and $\Delta\theta = |\theta_s - \theta_{xpt}|$, if the divertor plate is located right, the configuration is similar to a Snowflake minus. Again the designation adopted in Ref. 6 – "Asymmetric Snowflake minus" –is appropriate.

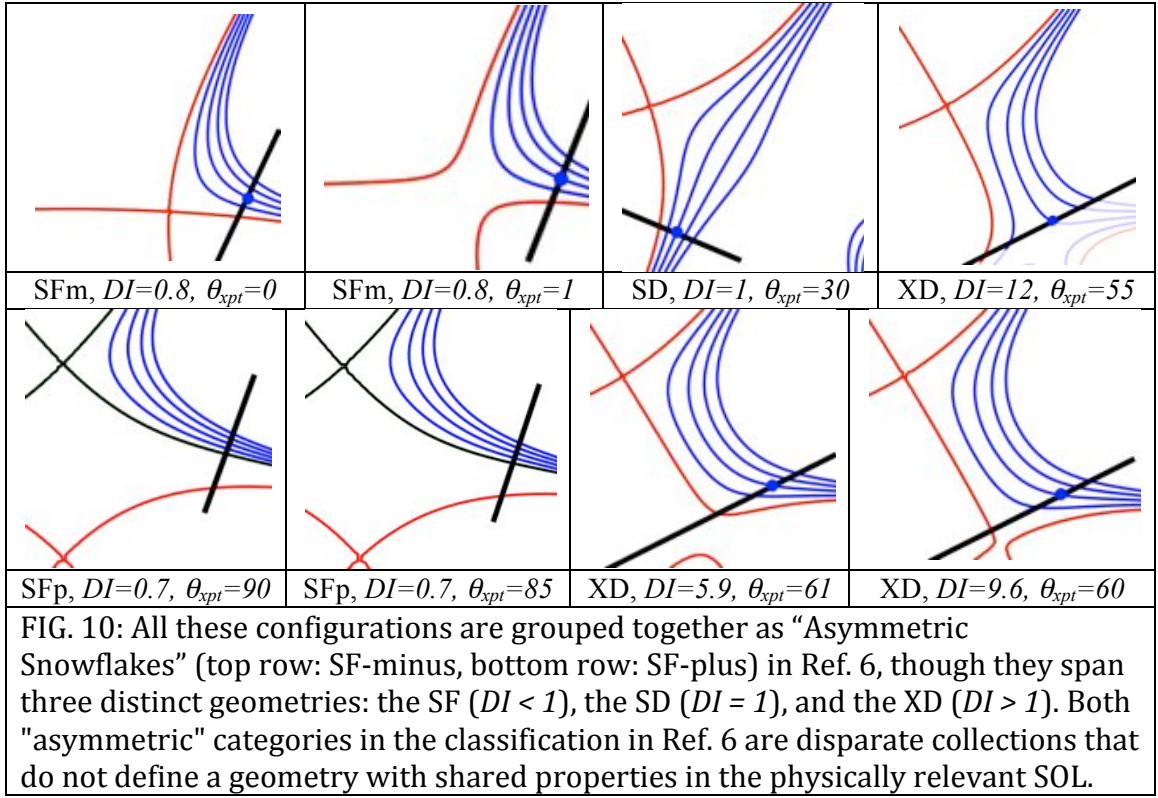

| SFm, *DI=0.8, $\theta_{xpt}$=0* | SFm, *DI=0.8, $\theta_{xpt}$=1* | SD, *DI=1, $\theta_{xpt}$=30* | XD, *DI=12, $\theta_{xpt}$=55* |
| SFp, *DI=0.7, $\theta_{xpt}$=90* | SFp, *DI=0.7, $\theta_{xpt}$=85* | XD, *DI=5.9, $\theta_{xpt}$=61* | XD, *DI=9.6, $\theta_{xpt}$=60* |

FIG. 10: All these configurations are grouped together as "Asymmetric Snowflakes" (top row: SF-minus, bottom row: SF-plus) in Ref. 6, though they span three distinct geometries: the SF (*DI < 1*), the SD (*DI = 1*), and the XD (*DI > 1*). Both "asymmetric" categories in the classification in Ref. 6 are disparate collections that do not define a geometry with shared properties in the physically relevant SOL.

3) For all $\theta_{xpt}$, as long as $\Delta\theta = |\theta_s - \theta_{xpt}|$ is sufficiently small, and $0 < d_s/d_{xpt} < 2$, the configuration is, *necessarily*, the same as the 2004 XD. Maximum flux expansion occurs at the plate by minimizing $\Delta\theta = |\theta_s - \theta_{xpt}|$ and bringing $d_s/d_{xpt}$ as close to unity as possible; this is precisely what the XD prescription was, and this is precisely what was done in the figures of the XD papers.
4) Though decreasing $d_{xpt}/a$ does lead to an absolute increase in flux expansion throughout the region, decreasing $\Delta\theta = |\theta_s - \theta_{xpt}|$ may be much more efficacious in increasing the flux expansion at the plate. As is evident from our analysis, *the flux expansion increases rapidly as the strike point ($d_s, \theta_s$) approaches the second X-point; for example, a factor of 10 can be gained by changing $|\theta_{xpt} - \theta_s|$ from 20 degrees to 2 degrees (Eq. 10) when $d_s=d_{xpt}$.* This is surely far more effective than modifying $d_{xpt}/a$ within the experimentally exploited range $0.2 < d_{xpt}/a < 0.4$.

The XD prescription, where the second X-point is downstream near the plate, $\left[(d_s,\theta_s) \to (d_{xpt},\theta_{xpt})\right]$, is the most effective way to create large (order of magnitude or more) increases in flux expansion at the plate. Since spreading heat at the divertor plate is one major mission of the advanced divertor enterprise, and it is the XD configuration that affects the largest flux expansion at the plate, the XD route is strongly indicated as the best divertor choice.

The categories "Asymmetric Snowflake minus" and "Asymmetric Snowflake plus" together fail to define a unified "class" of magnetic geometry for the exhaust SOL flux surfaces because this class would erroneously include the XDs; i.e., it would span physically distinct magnetic geometries (as distinguished by the visually obvious shape of the SOL flux surfaces, and by a variety of quantitative indices). This can be seen in Fig.10. A classification that defines a family based solely on ($d_{xpt},\theta_{xpt}$) without knowing ($d_s,\theta_s$) will not be very useful. The absence of the SOL and the divertor plate in the MagVAC analysis is a deficiency in its utility to capture the full essence of divertor physics.

### D. Non-uniqueness of the field outside the plasma

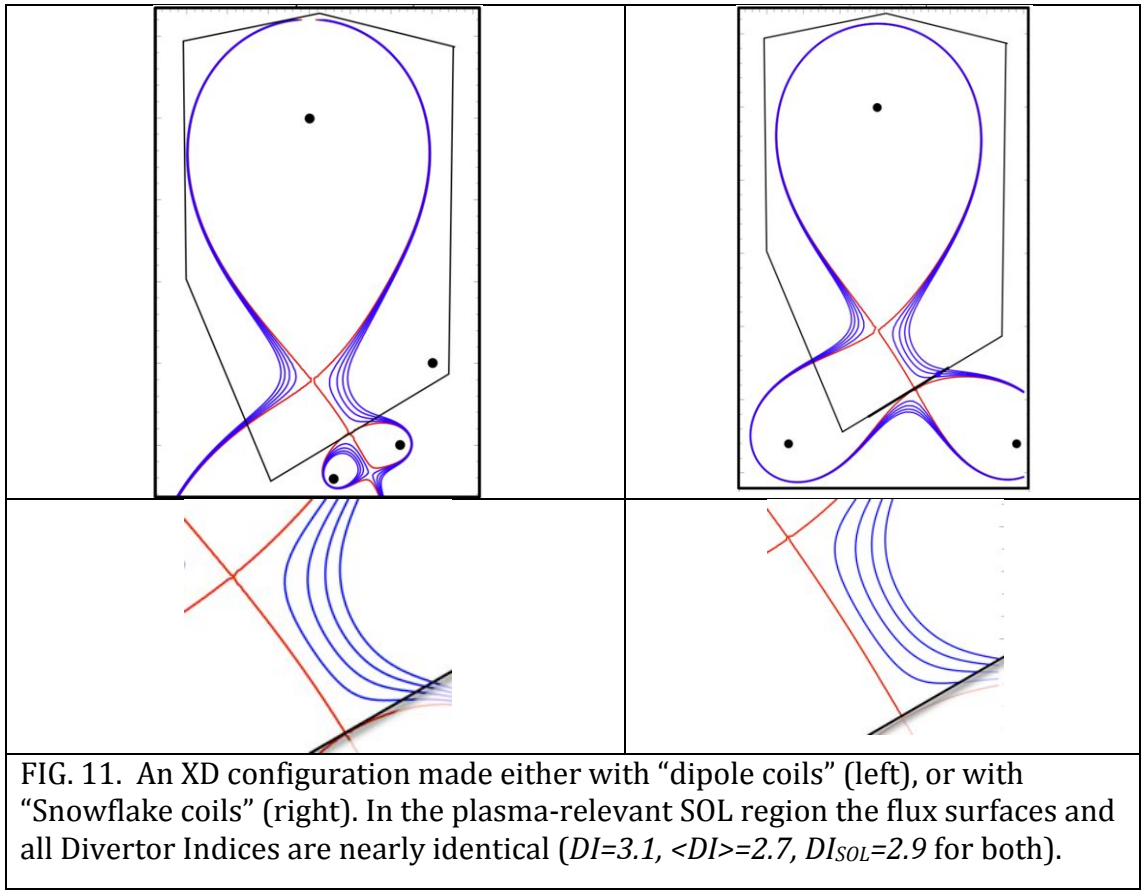

FIG. 11. An XD configuration made either with "dipole coils" (left), or with "Snowflake coils" (right). In the plasma-relevant SOL region the flux surfaces and all Divertor Indices are nearly identical (*DI=3.1, <DI>=2.7, DI$_{SOL}$=2.9* for both).

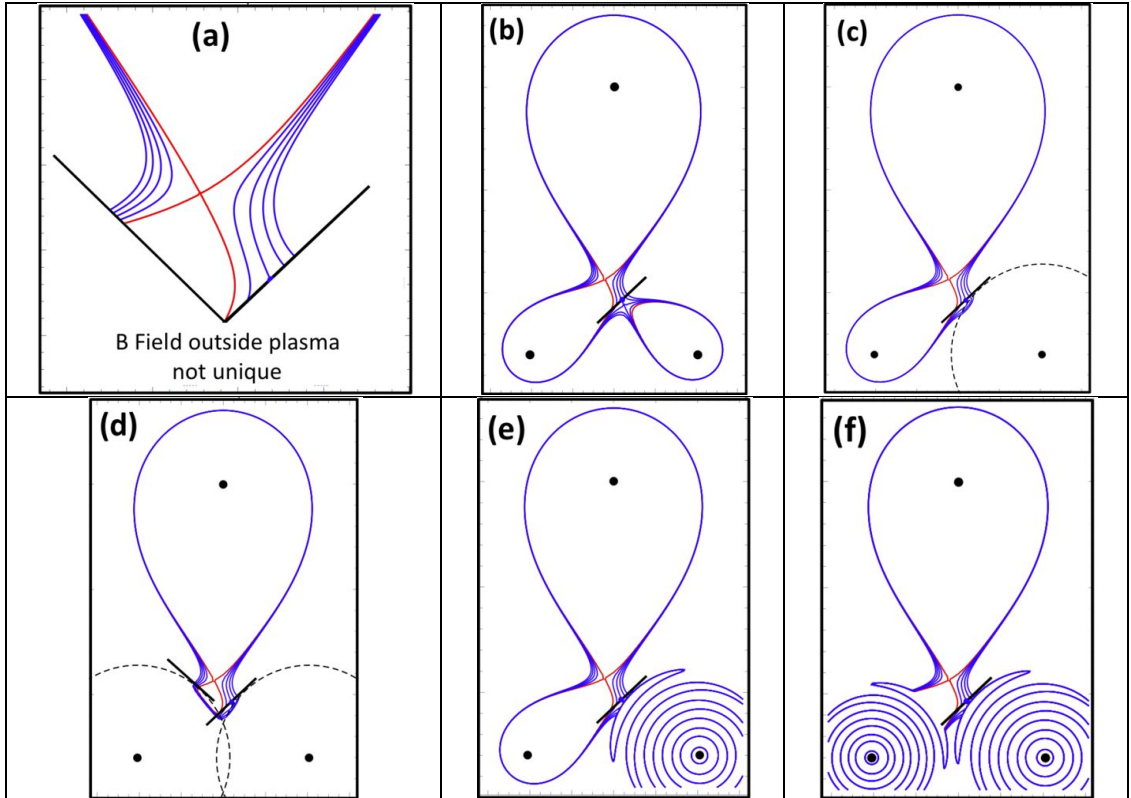

FIG. 12. The separatrix flux surfaces in the asymptotic region used by Ref. 6 to classify all two-X-point divertors as Snowflakes. In all cases the magnetic fields are *identically the same* in the region inside the plasma and SOL shown in (a). Case (b), produced as in Ref. 6, shows a three-lobe structure like a Snowflake in the region outside the plasma. In (c), the coil in the bottom right is replaced by a surface current on the dotted circle. In (d), both bottom coils are replaced. In (e) and (f), the finite radius coils have an oscillatory current density with the same net current. These finite radius coils only modify fields inside them, so cases (c)-(f) do not have the Snowflake appearance (no three-lobe flux surfaces in the asymptotic region), despite being identical to (a) in the plasma region. Hence, any classification based on fields outside plasma and SOL is not unique.

Another problem with the MagVAC classification is that it is non-unique: there exists an infinity of current distributions (and coils) outside the plasma region that can create exactly the same magnetic field inside the region. We illustrate this for the divertor geometries by creating practically indistinguishable SOL magnetic fields via various arrangements of wire currents, and currents in solid conductors (Figs. 11 and 12). The point is that the fields outside the plasma and SOL are rather different, but since all of them lead to same field in the SOL, the structure of the outside field has no relevance for the plasma exhaust controlled by the SOL field. Hence, shared properties in the region outside plasma (such as a snowflake-like structure in the asymptotic region away from the X-points) are unsuitable for defining a physically relevant class.

## E. Normalized x-point separation is not a distinguishing metric

An issue which we need to address before proceeding further concerns the perception that the normalized X-point separation $d_{xpt}/a$ could be used to differentiate between the XD and the SFD. The XD prescription never stated nor implied any restriction on the values of $d_{xpt}/a$. In the published work[1,2,3,4,5,10,11,12,13] (summarized in Table II); $d_{xpt}/a$ in XD examples (2004-2007) are not *materially different* from the Snowflake examples (2007-8), or recent experiments.

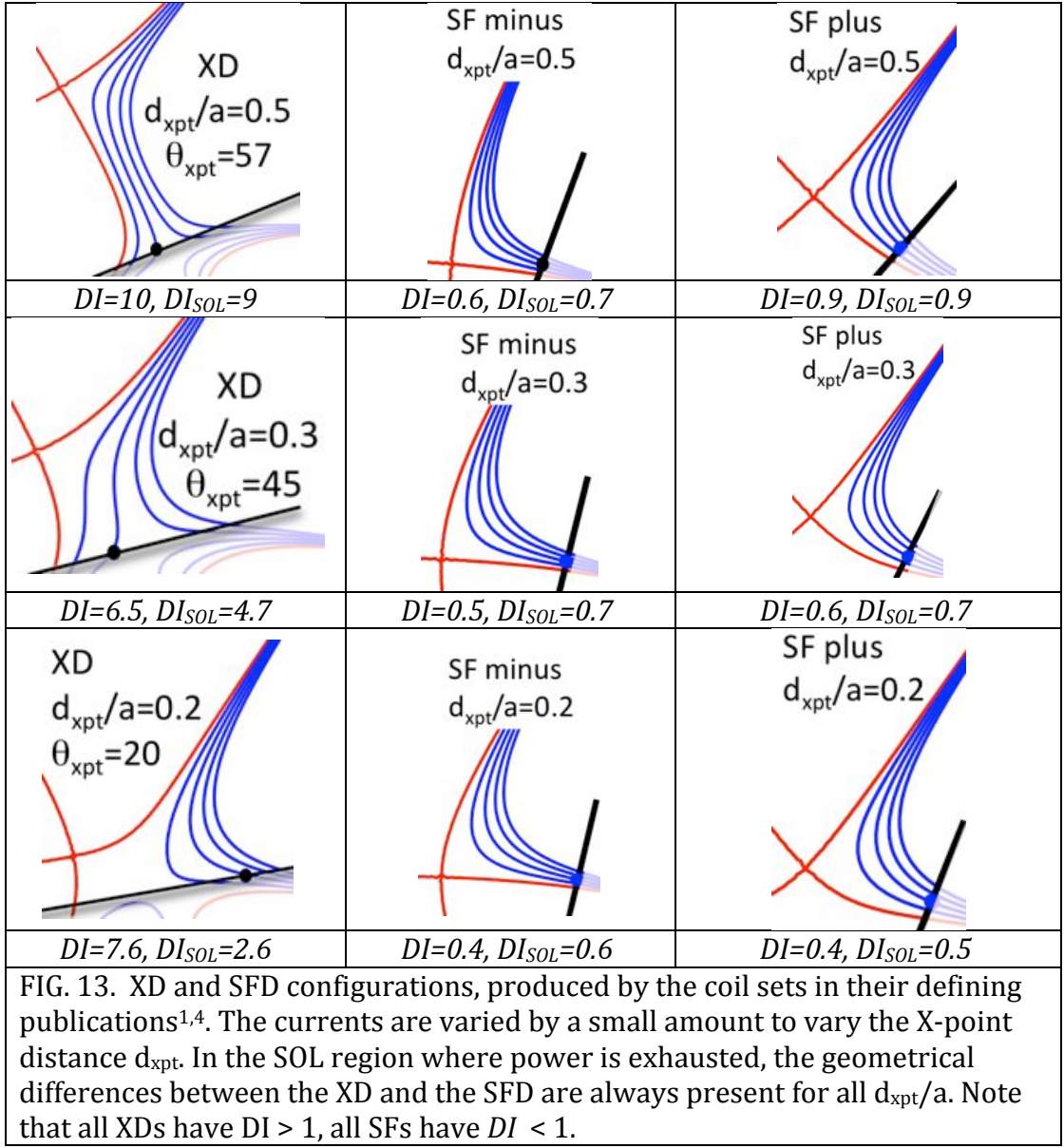

FIG. 13. XD and SFD configurations, produced by the coil sets in their defining publications[1,4]. The currents are varied by a small amount to vary the X-point distance $d_{xpt}$. In the SOL region where power is exhausted, the geometrical differences between the XD and the SFD are always present for all $d_{xpt}/a$. Note that all XDs have DI > 1, all SFs have *DI* < 1.

To fortify this point further, we show in Fig.13, a variety of XD and SFD cases with different but overlapping values of *$d_{xpt}/a$*; within the expected range, both geometries are quite robust to the change in X-point distance and retain their respective character.

| CASE | $d_{xpt}/a$ |
|---|---|
| 1) XD 2004[1]   NSTX | 0.53 |
| 2) XD 2004[1]   ITER | 0.62 |
| 3) XD 2004[1]   CREST | 1.00 |
| 4) XD 2007[3]   NSTX | 0.48 |
| 5) XD 2007[3]   ITER | 0.46 |
| 6) SFD plus[4,5,6] | 0.36 |
| 7) SFD minus[4,5,6] | 0.36 |
| 8) TCV   (experiment) SFDm | 0.44 |
| 9) NXTX  (experiment) 2009 | 0.23 |
| 10) NSTX  (experiment) 2010 | 0.41 |
| 11) DIII-D (experiment) 2012 | 0.26 |
| TABLE II. X-point distances normalized by the minor radius, $d_{xpt}/a$, for published advanced divertors.  Theoretical examples of both XDs and SFDs as well as experiments all contain cases with $d_{xpt}/a \sim 0.4$. | |

We conclude this Section by noting that the end results of the two classification schemes, MagSOL (this paper), and MagVAC[6] are strikingly different. The MagVAC, by relying on the non-unique field in the region outside the plasma, extends the SFD family to group together (and in the process subsume) physically distinct geometries like XD and SFD) while the MagSOL scheme, based on the physics of the SOL, tends to define and separate the domains of applicability of the XD and the SFD. Both XD and SFD are valid but independent magnetic solutions to the divertor exhaust problem. Both offer "legitimate" and easily defined alternatives to a Standard Divertor, albeit with different physical characteristics.

## IV. Divertor Geometry and Detachment Physics

Somewhat secure in the knowledge that we can not only produce but also possess necessary tools to describe, categorize, and label advanced divertors (in present and in future machines), we are ready to discuss the physics associated with these geometries. We begin with the physics of detachment.

### A. General considerations

At high levels of atomic dissipation (of power and momentum) the SOL plasma may manifest "detachment" from material surfaces. For the Standard Divertor , full detachment greatly reduces the heat flux on the divertor plate, and also greatly reduces the plasma temperature, resulting in a substantial reduction in plate erosion. Detached operation would be highly desirable as it will reduce the technological challenges of steady state power exhaust, but regrettably, experiments find that the main plasma suffers serious degradation for sufficiently strong detachment. Full detachment often results in a loss of loss of H-mode confinement quality[19,20,21,22], and at high density, can lead to a disruption[23].

A strong detachment front (initially formed near the plate) tends to move towards the main X-point bringing the cold plasma (sometimes termed an X-point MARFE) to the boundary of the main plasma. The presence of a cold, highly radiating plasma at the edge is suspected of causing deleterious effects on H-mode confinement, and on disruption likelihood. These drawbacks have, so far, prevented the fully detached regime from being considered a primary candidate for burning plasma devices, despite its other attractions.

Note that advanced divertors modify the magnetic field structure in the same region where the detachment front progresses from the divertor plate to the core X-point. If the new divetor geometries could enable fully detached operation without degrading the main plasma, the resulting benefits could be enormous. Here, we give several arguments to indicate that the XD and SFD affect the progression of the detachment front in opposite ways – the XD (SFD) retards (accelerates) the movement of the detachment front towards the main X-point. We note that these arguments were very briefly mentioned in the 2004 XD paper, but the details have not been presented till now.

For both the XD and SFD, there can be a substantial reduction of heat flux (compared to a Standard Divertor ) on the divertor plate. Essential differences in *plasma* behavior, however, arise when *additional* physics via neutral effects is brought into play. Without neutrals, for instance, the interaction of the plasma with the plate is described only by the sheath boundary condition. It is a surprising fact that this boundary condition is virtually insensitive to the angle of the magnetic field with the plate, and this is how flux expansion manifests in the sheath boundary condition[24]. Oddly enough, if neutrals are neglected, the cross-sectional area of the interaction with a material plate (increased by the larger flux expansion of advanced divertors) has almost no effect on the *plasma* (although it evidently will reduce heat flux to the plate). This remarkable physics "verity" could be stated in a more direct and telling manner: if a Standard Divertor had the same line length and the same SOL width as some XD or SFD, and if neutrals were neglected, the SOL plasma would be the same in all cases (as measured along a field line).

It is the interaction with neutrals that makes the "interaction area" affect the divertor function in a substantive way. According to Krasheninnikov[25,26], as detached conditions are approached, the plasma acquires a neutral gas "buffer" near the plate. Of course, close contact of the plasma with neutrals is accompanied by plasma energy losses – from ionization, charge exchange, enhanced atomic radiation, etc. One aspect of neutral dynamics implicit in the considerations of Ref 25 (and closely related arguments) is the expectation that a larger (smaller) interaction area *with neutrals* will result in larger (smaller) energy losses from the plasma[26]. Although the plasma behavior is determined by its dynamics along a magnetic field, the neutral dynamics is not. In an axisymmetric configuration, the cross-sectional area of interaction between the plasma and the neutral buffer depends on the shape of the plasma in the poloidal plane.

And since the "shape of the plasma in the poloidal plane" is almost synonymous with the magnetic geometry of the SOL, one expects that the divertor geometry will affect the dynamics of the detachment front (see Fig.14). The geometry will influence the dynamics by affecting energy losses in the plasma as neutrals penetrate into the plasma region. If the energy losses increase (decrease), the detachment front tends to move towards (away from) the heat source – the main plasma.

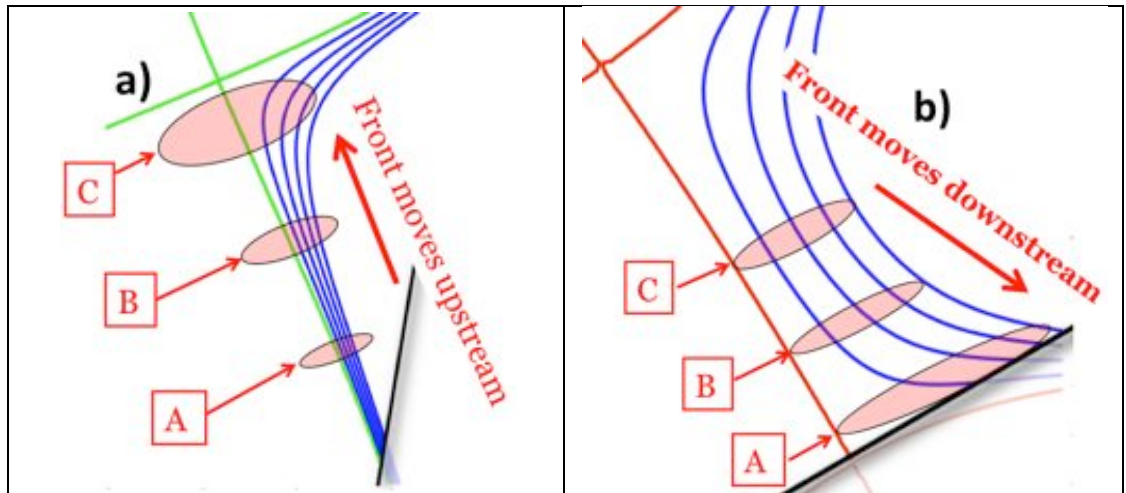

FIG. 14. The plasma-neutral interaction area of a) a Standard Divertor increases as the detachment front moves toward the main X-point. Thus, energy losses increase, leading to an unstable feedback, so that the front moves toward the core X-point. An XD geometry b) with flared field lines near the plate ($DI > 1$) reverses this feedback so the front could be arrested near the divertor plate.

Below, we describe the details of the three major mechanisms that affect the losses:
a) Variation in the contact area between the plasma and the neutral buffer
b) Variation in the upstream plasma pressure due to parallel thermal conduction
c) Geometrical effects on detachment

All of these effects are *directly* affected by the magnetic structure of the SOL in the exhaust region – and hence, we expect that categorizations based on that structure will be critical in correlating detachment behavior with divertor type. The net effect on detachment follows the same pattern for each of these mechanisms: relative to the Standard Divertor, the XD should have a much reduced tendency for the front to move upstream from the plate toward the core X-point, whereas for the SFD, there is an enhanced tendency to do so.

**B. Variation in the contact area between the plasma and the neutral buffer**

Because of the dependence of the plasma-neutral contact area on the local flux expansion, divergent field lines (the XD) and highly convergent field lines (SFD)(relative to SD) will trigger different feedback responses as the detachment front proceeds upstream from the divertor plate toward the main plasma X-point:

1) For a Standard Divertor , the flux expansion increases as the detachment front goes upstream. Consequently the contact area, and the associated energy losses increase. This positive feedback tends to cause a radiation collapse of the front so that it moves even further toward the main plasma X-point

2) The 2007 SFD geometry accentuates this tendency, since the flux expansion increases even more rapidly as the main plasma X-point is approached. Hence, the dynamics tend to make the detachment front proceed towards the main plasma X-point even more strongly. However, the enhanced flux expansion near the main plasma X-point creates a larger plasma region around the main plasma X-point; the enhanced plasma region might insulate the main plasma from the deleterious effects of a nearby detachment front.

3) The 2004 X-Divertor has a uniquely stabilizing feedback in the region of field line flaring, since the flux expansion decreases as the main plasma X-point is approached. The favorable magnetic geometry tends to localize detachment fronts in the region near the divertor plate retarding movement toward the main plasma X-point.

**C. Variation in the upstream plasma pressure due to parallel thermal conduction**

As a detachment front progresses upstream, the line length from the atomic region to the outboard mid-plane decreases. This is a stabilizing feedback, since the decreased line length reduces the upstream pressure, and hence, the amount of energy loss in the atomic region. *We note that, compared to a Standard Divertor , this feedback is much stronger for an XD, and much weaker for an* SFD.

The highly differentiated behavior follows from the fact that the line length is distributed differently for various configurations. For an XD, a considerably larger fraction of the line length comes from the region near the divertor plate. This is because the poloidal field is smaller near the plate, so that a field line travels a larger distance around the torus for a given increment of poloidal length. Conversely, for the SFD, a smaller fraction of the line length occurs from the region near the plate (since a large fraction arises from the region near the main plasma X-point). Hence, XD configurations will experience the strongest stabilizing feedback from this mechanism. The Standard Divertor  has a smaller fractional decrease. The Snowflake configurations have the smallest fractional reduction in line length, and hence, the weakest stabilizing feedback.

One way that variations in line length affect atomic dissipation is through the effect on upstream pressure. Variations in the upstream pressure also affect the energy losses, as indicated in Ref. 25. The neutral interaction is strong at a temperature (determined by atomic physics) roughly in the range 5-10 eV. The energy losses in that region are stronger if the plasma density is higher (which will also make the neutral density higher). The pressure is approximately constant along a field line up to the region of strong neutral interaction, and hence the plasma density in the neutral region is of order the upstream plasma pressure divided by the characteristic neutral temperature. The upstream pressure is known from the upstream density and temperature. We consider the effect of geometry on the temperature.

The upstream temperature is determined by parallel thermal plasma conduction long the field line. The relevant field line region is between the atomic region and the outboard mid-plane where the heat is conducted into the SOL from the main plasma. For a fixed power input and upstream SOL width, the upstream temperature will increase as the field line length increases. For example, in the two-point model where only parallel Spitzer thermal conductivity is considered, the upstream temperature varies as the 2/7 power of the line length. The upstream pressure has an important component from the ion temperature, determined by parallel transport as well. However, for typical parameters, this transport may be in a kinetic regime. In addition, electron-ion equilibration in the SOL can be an important factor for the upstream ion temperature. Despite the complexity of an actual SOL, all these dynamics have the feature that one expects a higher upstream pressure for a longer field line. In short, based on the plasma temperature, upstream pressure should be a monotonically increasing function of the line length.

### D. Geometry effects on detachment

For both the physical effects described so far, the stabilizing effect on the movement of the detachment front from upstream thermal conduction is largest for the X-Divertor, smallest for the SFD, and intermediate for the Standard Divertor .

This leads to the prediction that the X-divertor configuration (i.e., sufficiently high *DI* values) can attain higher levels of radiative dissipation than a standard divertor, *without suffering degradation of H-mode confinement*.

Of course, the actual detachment behavior is quite complex, with many factors at work simultaneously. The preceding arguments are indicative that one should expect significant differences in the three classes of divertor geometries: the geometries with convergent flux surfaces (like a Standard Divertor ), the geometries where field lines are even more rapidly convergent (SFD), and geometries with flux surfaces more slowly convergent, or divergent, near the plate (XD).

We find it very encouraging that preliminary experiments may be demonstrating behavior that is consistent with these basic physical arguments. For the Snowflake geometry, TCV finds the strongly radiating region close to the main plasma X-point. On the other hand, NSTX and DIII-D experiments, which have created geometries like the XD (flared near the divertor plate), find that the strongly radiating region stays near the divertor plate.

Edge simulations[27,28] have also demonstrated that radiation is enhanced in SFD geometries as compared to SD geometries, and detachment commences sooner. Same simulations also show that the SFD has a greater tendency for the radiation front to collapse to the core boundary. This is in general agreement with the expectations of this section.

We note that it has been speculated that the magnetic geometry can lead to enhanced plasma fluctuations[29] (due to increased $\beta_{poloidal}$ ) that can increase the SOL width. So far there has been no rigorous calculation of turbulent transport to theoretically support this speculation. To the best of our knowledge, there are no measurements of enhanced electromagnetic fluctuations dependent on the divertor geometry in the region of high SOL $\beta_{poloidal}$ that could lend experimental support to this conjecture. Hence, our discussion of detachment has been based solely on the physical processes whose importance is well accepted, i.e., the plasma-neutral interactions.

Finally, we end by noting an interesting proposal called an "X-point Target" divertor[30], where a second X-point is located downstream but is still in the SOL region; i.e., on the plasma side of the divertor plate. It is argued[28] that this will assist in maintaining stable full detachment. The "X-point Target" concept fits within the definition of the XD in the abstract in the X-Divertor paper[1], and relies on detachment behavior that is similar to that predicted[1]. The arguments of this section indicate that the "X-point Target" concept has considerable merit, and further exploration is highly desirable (theoretically, via simulations, and in experiments). According to the arguments here, the second X-point would act as an "attractor" for the detachment front – a front located either upstream or downstream of the second X-point would tend to move closer to that point. Hence, such a placement of a second X-point might very well further assist in maintaining full detachment while keeping the front away from the main X-point. Such a concept might also be beneficially applied to the Super-X Divertor[7,8,9].

## V. Extrapolation to Future machines

The considerable initial success of TCV-NSTX-DIIID experiments in implementing advanced divertor geometries is a striking achievement in fusion physics and engineering. One must, then, ask if the realized geometries could be replicated in future machines (ITER/reactor). And if yes, how effective would they be in solving the very challenging exhaust problem such devices will have to face. Since the divertor action is controlled by the interaction of the magnetic geometry with the SOL, it is essential to understand the similarities and differences in the character and shape of the SOL flux surfaces peculiar to current and future machines.

Recent projections of SOL width ($\Delta$) for ITER and reactors imply that the SOL power width becomes much thinner as we progress from present experiments to such burning plasmas[16,17]. In fact, the projection of a smaller $\Delta$ has provided a very strong additional motivation for examining novel/advanced divertor configurations: the dimensionless ratio, $\Delta/a$ (=the SOL width divided by the minor radius) for a reactor or ITER is an order of magnitude smaller as compared to the present experiments. We must, therefore, exercise extreme caution in developing a dependable methodology for extrapolating from present experiments to a reactor.

The MagSOL analysis (identifying physical metrics and categories) given in Sec. III, based on the properties of the magnetic field in the region of the plasma SOL, can be a key to a dependable extrapolation. In fact, to extrapolate, confidently, to plasmas for which $\Delta/a$ could be different by an order of magnitude, correct formulation of such measures is an imperative.

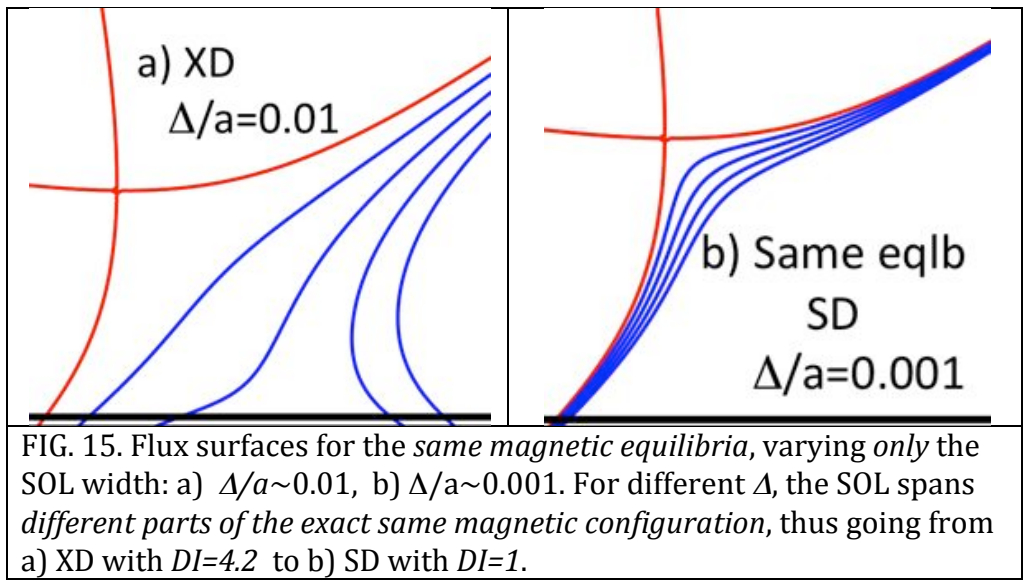

FIG. 15. Flux surfaces for the *same magnetic equilibria*, varying *only* the SOL width: a) $\Delta/a \sim 0.01$, b) $\Delta/a \sim 0.001$. For different $\Delta$, the SOL spans *different parts of the exact same magnetic configuration*, thus going from a) XD with *DI=4.2* to b) SD with *DI=1*.

We present below an example of the difficulties that can arise from incorrect classification. We consider a magnetic geometry that is similar on ITER and on current experiments in a *gross* sense (e.g. similar according to MagVAC). However, it should have a different divertor behavior in the two devices because their SOLs occupy different regions of space *in the exhaust SOL region* where the magnetic fields are quite different. Specifically, in this section, we will see that a MagVAC categorization based on the angle $\theta_{xpt}$ could lead to extrapolation errors. This is because the SOL is missing in MagVAC, and the SOL geometry is determined by the angular separation $\Delta\theta = |\theta_s - \theta_{xpt}|$ and not by $\theta_{xpt}$ alone. So unless the SOL, the SOL width, and the position of the wetted area on the divertor plate, were all a part of the analysis (like in MagSOL), no physically meaningful extrapolation is possible.

Things change drastically as we transition to ITER/reactor. Using the recent scaling given by Goldston[16,17], $\Delta/a$ is almost ten times smaller for an ST reactor (for instance, the ARIES ST[31]) as compared to NSTX. For ITER, $\Delta/a \sim 0.0005$-$0.001$, is also about an order of magnitude less than NSTX. To get an idea about ITER/reactor relevance of the current experiments, we plot in Fig.15 the SOL from the same magnetic field as for the case with $\Delta/a = 0.01$, *but assume a reactor relevant $\Delta/a = 0.001$*. Note that for the identical magnetic geometry, the magnetic field in the SOL looks extremely different. The flaring is substantially reduced – the SOL experiences reduced flux expansion; the ratio of the footprint on the divertor upstream SOL width is only ~20, and the Divertor Indices are approximately = 1, like an SD.

For the reactor-relevant SOL, residing in an NSTX-like geometry, the flux expansion is greatly reduced, to a value more typical of a Standard Divertor . The same magnetic geometry produces much less benefit for a much smaller SOL width.

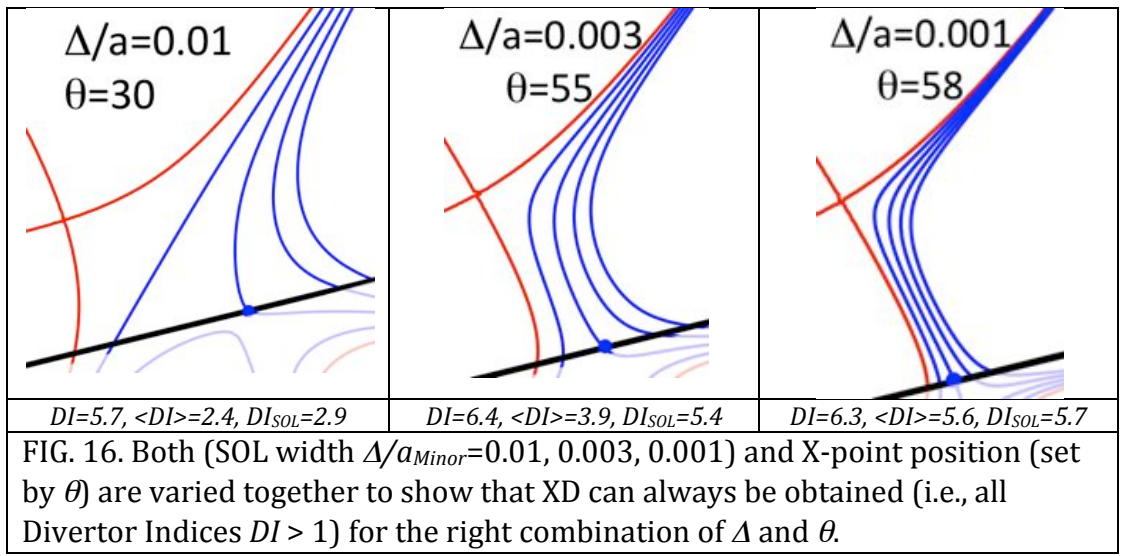

| $DI$=5.7, $<DI>$=2.4, $DI_{SOL}$=2.9 | $DI$=6.4, $<DI>$=3.9, $DI_{SOL}$=5.4 | $DI$=6.3, $<DI>$=5.6, $DI_{SOL}$=5.7 |

FIG. 16. Both (SOL width $\Delta/a_{Minor}$=0.01, 0.003, 0.001) and X-point position (set by $\theta$) are varied together to show that XD can always be obtained (i.e., all Divertor Indices $DI > 1$) for the right combination of $\Delta$ and $\theta$.

It is worth noting that the NSTX experiments did exactly what was needed to implement an X-Divertor on NSTX – they placed the X-point downstream, and within the SOL footprint on the divertor plate. By doing so, they imparted flux expansion and longer line length to as large a fraction of the SOL as possible. The XD recipe, articulated in the original 2004 XD paper[1], and successfully implemented on NSTX[11,12], offers for ITER an explicit prescription for duplicating the same desirable physical effect: let the position of the X-point be dictated by the appropriate ITER SOL width; place it near the SOL footprint on the divertor. Such a procedure ensures that the geometry of the physically relevant plasma SOL (where power is exhausted) becomes similar on ITER and on NSTX leading to similar physical behavior.

To repeat the lesson of the last paragraph: for a reactor-like SOL width, the X-point must be suitably shifted to correspond to the position of the new, much smaller footprint on the divertor plate, compared to magnetic geometries optimized for $\Delta/a$ characteristic of present experiments.

The implementation of the preceding recipe is illustrated in Fig.16. Note that the picture is then similar to the 2004 X-Divertor for NSTX (Fig.2a).

This lesson is simply an expression of the obvious expectation that, if the geometry of the physically relevant plasma SOL is similar, for example, on ITER and on NSTX, physical behavior of these configurations will also be similar. Therefore, any index, used to characterize the geometry must also pass this test; similar values of the index must imply similar physical properties. For example, since the same angular difference $\Delta\theta = |\theta_s - \theta_{xpt}|$ produces the same physical effect (Sec. IIIc), when the strike point angle changes, the X-point angle must commensurately change to compensate.

As a corollary, the angular location $\theta_{xpt}$ of the second X-point, by itself, is not enough to determine physics; a geometry with a given $\theta_{xpt}$, while bringing substantial benefits to the NSTX plasma, may yield significantly less beneficial effect for a reactor with much smaller SOL width.

To make sure that the current experiments do serve as a useful guide for the future, we must interpret the experiments in terms of metrics and concepts that focus on the factors that control divertor action – the physics of the divertor exhaust. It is the structure of the magnetic field in the SOL that provides the relevant magnetic input to the divertor physics. And that is precisely the basis of the MagSOL classification proposed in this paper.

## VI. Conclusions:

Through detailed arguments and invoking analytical and numerical tools, we have attempted to put in perspective recent progress in theoretical investigations and experimental implementation of the Advanced Divertors with two X-points. Such divertors may very well be necessary for solving the enormous heat exhaust problem that future fusion machines are likely to be saddled with.

The primary motivation of the paper was to understand the magnetic structure of these geometries so that we can better understand divertor physics and acquire dependable capabilities of extrapolation of the current experimental achievements to future machines. Realizing that divertor physics, to a large extent, is controlled by the magnetic structure in the SOL plasma, we introduced the SOL explicitly into the magnetic field analysis. The resulting MagSOL formulation, obtained by a variety of conceptual arguments and numerical work, and fortified by a detailed analysis of the model analytical magnetic field (Ref. 6), is the main contribution of this work.

A major outcome of this long and in-depth MagSOL investigation is the formulation of what we call the Divertor Index *DI* (in several incarnations), a numerical index that differentiates between three basic divertor geometries; the Standard Divertor SD, and the two main advanced geometries (that employ only poloidal flux expansion): the 2004 X-Divertor and the Snowflake. The Standard Divertor SD (*DI = 1*) serves as the natural baseline, the SFD (*DI < 1*) and XD (*DI > 1*) depart from it in diametrically opposite directions. Plots of *DI* (Fig.7), obtained from numerical CORSICA equilibria, are in full agreement with analytical model calculations. In the SOL near the strike plate, the SFD flux surfaces are more convergent than the SD, while the XD flux surfaces are less convergent, in fact they can be divergent. We show why one expects flared XD SOLs with *DI > 1* to be more effective in reducing heat flux problems, and perhaps even allowing fully detached divertor operation, without degrading H-mode confinement.

The MagSOL analysis acquires particular relevance in the light of recent experiments in 2008-2012 (TCV[10], NSTX[11,12], DIIID[13]) that have translated the theoretical ideas from 2004 onwards into actual advanced divertors. In this context, the theoretical investigations of Ryutov *et al* [4-6,27-29] have played an important role.

According to the quantitative metrics developed in the MagSOL, recent NSTX and DIIID experiments are X-Divertors, while the TCV experiments are Snowflakes. Since different magnetics with relationship to the SOL imply correspondingly different physics, particularly with respect to the detachment dynamics, and the ability to extrapolate the current experimental results to ITER/reactor, the conceptual framework of MagSOL could augment the excellent experimental programs looking for an eventual magnetic divertor solution for fusion reactors.

## Acknowledgments

We gratefully acknowledge many colleagues, in particular, Prof. F. Waelbroek and the anonymous referee for insightful comments and suggestions. We also thank Dr. L. Lodestro for help with CORSICA. This work was supported by US-DOE grants DE-FG02-04ER54742 and DE-FG02-04ER54754.